\documentclass{article}
\usepackage{arxiv}

\usepackage[utf8]{inputenc} % allow utf-8 input
\usepackage[T1]{fontenc}    % use 8-bit T1 fonts
\usepackage{hyperref}       % hyperlinks
\usepackage{url}            % simple URL typesetting
\usepackage{booktabs}       % professional-quality tables
\usepackage{amsfonts}       % blackboard math symbols
\usepackage{nicefrac}       % compact symbols for 1/2, etc.
\usepackage{microtype}      % microtypography
\usepackage[numbers]{natbib}
\usepackage{doi}

\usepackage{graphicx}  % Enables inclusion of images
\usepackage{dcolumn}% Align table columns on decimal point
\usepackage{caption}  % Customization of figure and table captions
\usepackage{subcaption}  % Allows multiple subfigures with independent captions

\usepackage{multirow}  % Merges multiple rows in a table
\usepackage{booktabs}

\usepackage{bm}% bold math
\usepackage{amsmath}
\usepackage{amssymb}

\usepackage{xcolor}  % Enables colored text
\usepackage{color, colortbl}  % Adds color to tables and text
\definecolor{green}{RGB}{0,100,0} % For good improvements
\definecolor{red}{RGB}{200,0,0} 
\usepackage{hyperref}
\usepackage{cleveref}       % smart cross-referencing

\title{VIVALDy: A Hybrid Generative Reduced-Order Model for Turbulent Flows, Applied to Vortex-Induced Vibrations}

\date{}

\newif\ifuniqueAffiliation
\ifuniqueAffiliation % Standard variant of author block
\author{ \href{https://orcid.org/0009-0003-0164-5601}{\includegraphics[scale=0.06]{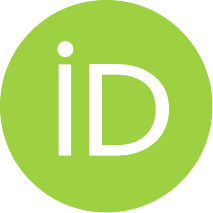}\hspace{1mm}Niccolò Tonioni}\thanks{https://niccolotonioni.github.io/} \\
	Pprime Institute,\\
	CNRS, Université de Poitiers, ISAE-ENSMA,\\
    Chasseneuil-du-Poitou, 86360, France.\\
	\texttt{niccolo.tonioni@univ-poitiers.fr} \\
	\And
	Lionel Agostini \\
	Pprime Institute,\\
	CNRS, Université de Poitiers, ISAE-ENSMA,\\
    Chasseneuil-du-Poitou, 86360, France.\\
	\And
	Franck Kerhervé \\
	Pprime Institute,\\
	CNRS, Université de Poitiers, ISAE-ENSMA,\\
    Chasseneuil-du-Poitou, 86360, France.\\
	\And
	Laurent Cordier \\
	Pprime Institute,\\
	CNRS, Université de Poitiers, ISAE-ENSMA,\\
    Chasseneuil-du-Poitou, 86360, France.\\
	\And
	Ricardo Vinuesa\\
	Department of Aerospace Engineering,\\
	University of Michigan,\\ 
    Ann Arbor, MI 48109, United States\\
    and\\
    FLOW, Engineering Mechanics,\\
    KTH Royal Institute of Technology,\\
    Stockholm, SE-100 44, Sweden.
}
\else
\usepackage{authblk}

\newbox{\orcid}\sbox{\orcid}{\includegraphics[scale=0.06]{orcid.pdf}} 
\author[1]{%
	\href{https://orcid.org/0009-0003-0164-5601}{\usebox{\orcid}\hspace{1mm}Niccolò Tonioni\thanks{\href{https://niccolotonioni.github.io}{niccolotonioni.github.io},\hspace{1mm}\texttt{niccolo.tonioni@univ-poitiers.fr}}}%
}
\author[1]{Lionel Agostini\thanks{\texttt{lionel.agostini@cnrs.fr}}}
\author[1]{Franck Kerhervé}
\author[1]{Laurent Cordier}
\author[2,3]{Ricardo Vinuesa\thanks{\href{https://www.vinuesalab.com/}{vinuesalab},\hspace{1mm}\texttt{rvinuesa@umich.edu}}}

\affil[1]{Pprime Institute, CNRS, Université de Poitiers, ISAE-ENSMA,Chasseneuil-du-Poitou, 86360, France}
\affil[2]{Department of Aerospace Engineering,University of Michigan, Ann Arbor, MI 48109, United States}
\affil[3]{FLOW, Engineering Mechanics, KTH Royal Institute of Technology, Stockholm, SE-100 44, Sweden}
\fi

\renewcommand{\headeright}{Preprint -- \textit{Phys. Rev. Fluids} \textbf{11}, 044902}
\renewcommand{\undertitle}{%
    This is a preprint of the article published in \textit{Physical Review Fluids}. \\
    The version of record is available at: \href{https://doi.org/10.1103/t7d2-mv5c}{https://doi.org/10.1103/t7d2-mv5c}
}
\renewcommand{\shorttitle}{VIVALDy}

\hypersetup{
pdftitle={VIVALDy: A Hybrid Generative Reduced-Order Model for Turbulent Flows, Applied to Vortex-Induced Vibrations},
pdfsubject={q-bio.NC, q-bio.QM},
pdfauthor={Niccolò Tonioni, Lionel Agostini, Franck Kerhervé, Laurent Cordier, Ricardo Vinuesa},
pdfkeywords={Reduced-Order Models, Turbulent Flows, Vortex-Induced Vibrations, Variational Autoencoders, Generative Adversarial Networks, Transformers},
}

\begin{document}
\maketitle

\begin{abstract}
	Developing reduced-order models applicable to fluid-dynamics problems involving complex geometries and different flow conditions remains a critical challenge for turbulent flows. This study introduces VIVALDy, a novel machine-learning framework that employs a hybrid $\beta$-Variational Autoencoder-Generative Adversarial Network ($\beta$-VAE-GAN) architecture with masked convolutions to extract dominant flow features into a compact latent space while preserving fidelity at solid-fluid interfaces. A bidirectional transformer then models the temporal evolution of these features, learning to predict flow trajectories from minimal sensor inputs. This two-stage approach enables the transformer to map sensor measurements to dominant flow variables identified by the autoencoder, advancing reduced-order modeling capabilities for real-time flow prediction. The effectiveness of the framework is demonstrated through application to a problem relevant to vortex-induced vibration (VIV) energy harvesting systems, reconstructing the turbulent flow around a one-degree-of-freedom moving cylinder. Validated against experimental data spanning fluid-structure interaction regimes of interest, VIVALDy accurately predicts different flow states using only the cylinder displacement. The framework demonstrates adequate performance in both reconstruction accuracy and statistical fidelity across diverse operating conditions, enabling efficient prediction of the turbulent flow phenomena governing vortex-induced vibration.
\end{abstract}

% keywords can be removed

% ----------------- Introduction
\section{Introduction}
\label{sec:01_introduction}
The increasing concerns about the global climate crisis and the United Nations Sustainable Development Goal 7 (\emph{ensuring access to affordable and clean energy}) have stimulated significant interest among public institutions in renewable energy-harvesting technologies \cite{WANG2020114902}. While the wind energy sector has mature technologies, wave and tidal energy remain in the proof-of-concept stage, limiting exploitation of Earth's vast water current resources \cite{UNEPFI_TurningTide_2021}. These technologies face unique challenges in marine and fluvial ecosystems: devices must maintain efficiency under varying flow conditions, and be unobtrusive with low maintenance costs due to biofouling. Vortex-induced vibration (VIV) systems offer a promising solution due to their simple structure and operation mechanism \cite{bernitsas2008vivace, schmider2024improved}.

VIV is a form of self-excited fluid-structure interaction (FSI) characteristic of bluff bodies, such as cylinders, immersed in a fluid flow \cite{annurev.fluid.36.050802.122128}.  VIV energy harvesting systems exploit these structural vibrations rather than suppress them, converting mechanical motion into usable energy. However, to maximize energy extraction from these devices under all operating conditions, real-time control strategies and rapid design optimization are essential. Therefore, there is a need for the development of Reduced-Order Models (ROMs) that can overcome the prohibitive computational costs associated with predicting the turbulent flow fields that govern VIV phenomena.

ROMs are techniques designed to create computationally efficient surrogate models by constructing low-dimensional approximations preserving the most informative features of the original system \cite{doi:10.1137/1.9781611977257}. Historically, research on ROMs has focused on linear methods with two main classes emerging: projection-based ROM (PB-ROM) and inference-based ROM (IB-ROM). PB-ROMs are inherently intrusive as they require explicit basis expansions and projections of the full-order model operators \cite{Cordier_Bergmann_2008_IVK_POD_Overview, Cordier_Bergmann_2008_IVK_POD_Applications}. In contrast, IB-ROMs offer a non-intrusive alternative. They infer surrogate operators directly from the data based on prior knowledge of governing equation structure, without requiring access to full-order operators \cite{schmid2010dynamic,annurev-fluid-121021-025220}. However, they may require regularization techniques as they are based on solving least-squares problems, which can be ill-conditioned and sensitive to noise \cite{mcquarrie2021data}. Furthermore, constructing full-order operators remains computationally expensive, necessitating data projections with associated truncation errors. Non-linear IB-ROMs have been developed to reduce truncation errors and improve accuracy \cite{peherstorfer2016data, kramer2024learning}. However, approximating both linear and quadratic operators of the Navier-Stokes equations may incur even higher computational costs. Another limitation of these models is their implementation, which is traditionally in a non-parametric form.  Consequently, they demonstrate an inability to generalize to variations in flow conditions, such as changes in incoming flow velocity a particular area of interest in the present study. Although some studies have explored how to extend these classical models to parameterized dynamic problems \cite{benner2015survey, mcquarrie2023nonintrusive}, this area has seen only limited development.

To address these limitations, researchers have increasingly explored machine-learning techniques for ROM development. The objective is developing non-linear approaches that improve accuracy, efficiency, and generalizability across flow configurations \cite{annurev-fluid-010719-060214,vinuesa2022enhancing}. Autoencoders (AEs) have gained particular attention due to their success in areas such as image compression. An AE, first introduced by Hinton and Salakhutdinov \cite{HinSal2006autoencoder}, is a neural network architecture comprising two primary (non-linear) mappings: an encoder and a decoder. The encoder compresses high-dimensional input data into lower-dimensional latent representations, also termed code. Subsequently, the decoder reconstructs the input data from this latent space, returning to the original high-dimensional space. The method is purely data-driven, with the network weights optimized by minimizing an objective function. Consequently, Autoencoders offer a non-intrusive, equation-free framework for constructing models that generalize across diverse flow parameters, provided these parameters are represented in the training dataset.

Agostini explored the application of AE for modeling a two-dimensional laminar cylinder wake flow \cite{agostini2020exploration}. In the study, the author compares the results obtained using Proper Orthogonal Decomposition (POD), a classical PB-ROM approach, with those obtained using AE for a latent space of dimension three. The results indicate that the AE surpasses POD in terms of reconstruction quality. However, a limitation of AE is the lack of orthogonality of the low-dimensional representations. Inspired by the Cluster-ROM algorithm \cite{kaiser2014cluster}, Agostini also constructed a probabilistic dynamical model in the AE latent space using spectral clustering and Markov chains. This statistical model allowed the author to quantify the likelihood of specific clusters occurring under given flow states. Since each cluster corresponds to distinct dynamics within the latent space and is associated to specific flow states, this approach provided physical interpretability of the underlying flow behavior.

Eivazi \textit{et al.} \cite{eivazi2022} extended this line of search to Variational Autoencoders (VAEs), which generalize AEs through stochastic mappings for low-dimensional representation of flow dynamics. Specifically, the authors employed a $\beta$-VAE \cite{higgins2017beta}, a modified VAE architecture that introduces a hyperparameter $\beta$ to balance reconstruction fidelity against latent space regularization toward a standard normal distribution. This variation enables the user to control the model's training to prioritize \emph{disentanglement} in the latent space, i.e., to separate independent factors of variation in the data. Eivazi \textit{et al.} \cite{eivazi2022} applied a $\beta$-VAE  to flows in a simplified urban environment, and compared the results against AE and POD. Their results demonstrate that the $\beta$-VAE not only achieves better reconstruction of the flow fields for a given compression ratio, but it also achieves more orthogonal latent variables compared to AE. 

In a series of subsequent papers, Wang \textit{et al.} \cite{WANG2024109254} and Solera-Rico \textit{et al.} \cite{solera2024beta} combined $\beta$-VAEs with other neural architectures to model the temporal dynamics in the latent space and develop ML-driven ROMs for predicting turbulent flows. Specifically, they compared the performance of Long Short-Term Memory (LSTM) \cite{hochreiter1997long}, a recurrent architecture introduced for learning long-range dependencies in sequence data, with Transformers \cite{vaswani2017attention}, which, in contrast, exploit self-attention mechanisms to model dependencies without relying on recurrence, offering advantages in parallelization. Both studies concluded that Transformers have lower errors and the best long-term reconstruction performance.

\begin{figure}
\centering
\includegraphics[width=.9\linewidth, trim=0cm 5cm 0cm 5cm, clip]{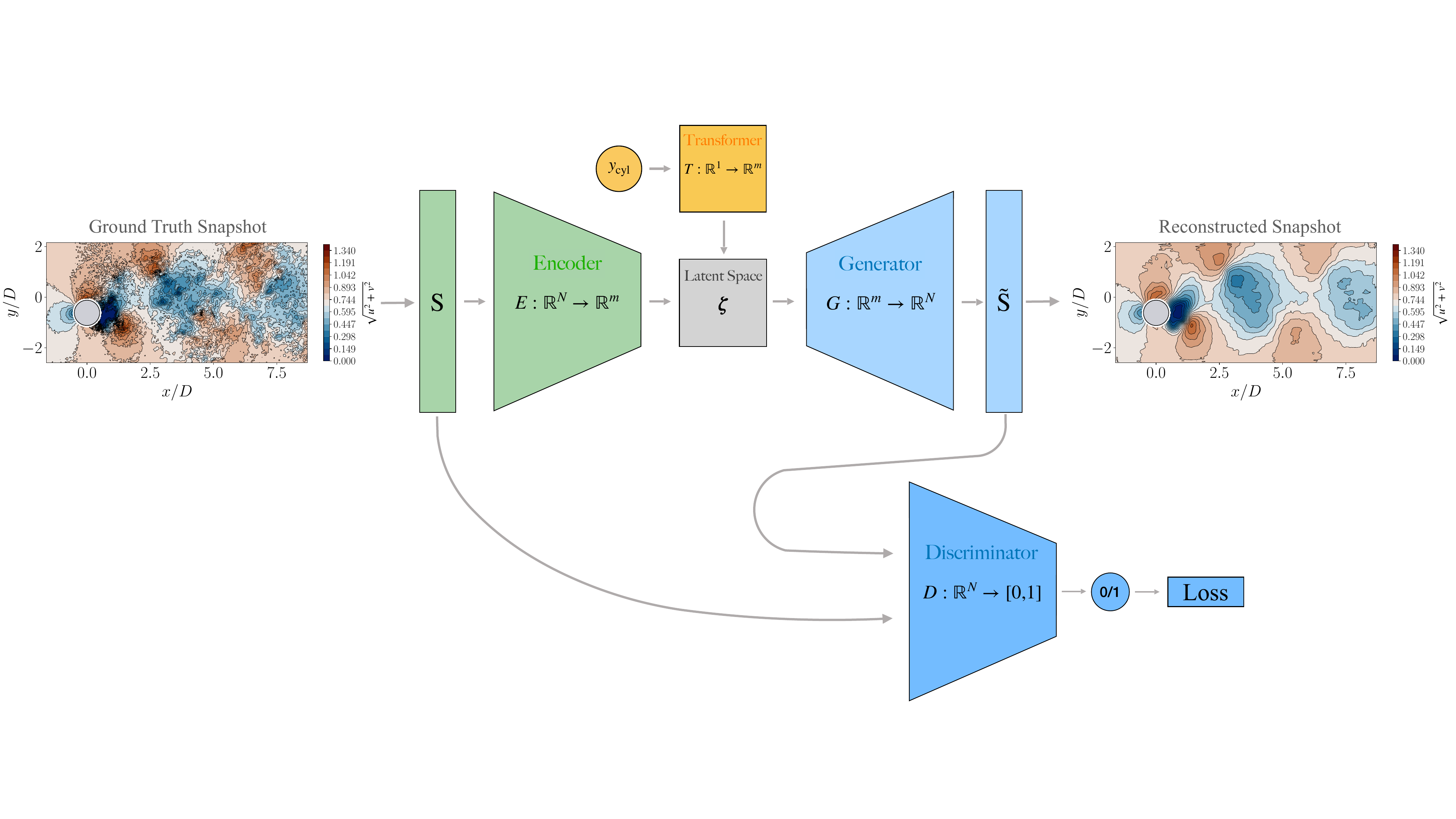}
\caption{\protect Schematic representation of VIVALDy framework. \textit{Training phase:} The $\beta$-variational autoencoder ($\beta$-VAE) and discriminator are trained simultaneously in a generative adversarial framework, where the decoder serves as generator and receives evaluative feedback from the discriminator. A transformer model is then trained to predict latent variable evolution using only cylinder displacement $y_{\mathrm{cyl}}$ as input. \textit{Inference phase:} Only the transformer and decoder are retained to generate flow field predictions from displacement signals.}
\label{fig:intro/data-driven-framework}
\end{figure}

Inspired by previous works, this paper introduces VIVALDy (Vortex Induced Vibration Autoencoder for Low-dimensional Dynamics). While named after its initial application to VIV systems, this method constitutes a general-purpose, data-driven ROM (see Figure \ref{fig:intro/data-driven-framework} for its overall structure). In contrast to existing methodologies, VIVALDy can reconstruct the flow state across varying flow conditions from minimal sensor measurments. When applied against a newly acquired experimental dataset encompassing a range of fluid-structure interaction (FSI) regimes directly relevant to practical VIV energy-harvesting applications, VIVALDy successfully reconstructs the turbulent flow around a one-degree-of-freedom cylinder using only the cylinder displacement as input. Moreover, the model's predictions are both computationally and memory-efficient as it operates directly in the latent space.

From a physical point of view, VIVALDy's approach is feasible due to the coupling between the cylinder's motion and the surrounding flow field, resulting in correlated signals. From a computational perspective, this is made possible by combining a novel hybrid generative architecture for latent-feature extraction with a bidirectional transformer designed to learn the non-linear correlations between the cylinder kinematics and the flow dynamics. The hybrid architecture, which is based on $\beta$-VAE and Generative Adversarial Networks (GANs) \cite{goodfellow2014generative}, is motivated by recent works in image compression \cite{tschannen2018deep, agustsson2019generative, mentzer2020high}. These works demonstrate that this combination leverages the individual strengths of both models, resulting in improved distribution-preserving properties compared to a standard VAE and a more structured, informative latent space than that offered by a standalone GAN. 

The paper is organised as follows: Section \ref{sec:02_dataset} details the experimental dataset and its preparation. Section \ref{sec:03_methods} provides a detailed description of the VIVALDy framework. Section \ref{sec:04_results} presents the principal results, which are discussed in Section \ref{sec:05_discussion}. Finally, Section \ref{sec:06_conclusions} summarizes the main contributions and discusses future outlook.

% ----------------- Dataset

\section{Dataset Generation and Preparation}
\label{sec:02_dataset}
This section details the experimental dataset employed for the training and validation of VIVALDy. The dataset comprises streamwise and crosswise velocity fields $(u, v)$ of an elastically mounted cylinder undergoing vortex-induced vibrations in the cross-flow direction. The data were acquired using time-resolved Particle Image Velocimetry (PIV). The following subsections describe the experimental setup and the preparation of the acquired flow snapshots for model input.

\subsection{Experimental Setup and Data Acquisition}
\begin{figure}[h]
\centering
\begin{minipage}[b]{0.35\textwidth}
    \centering
    \includegraphics[width=\linewidth]{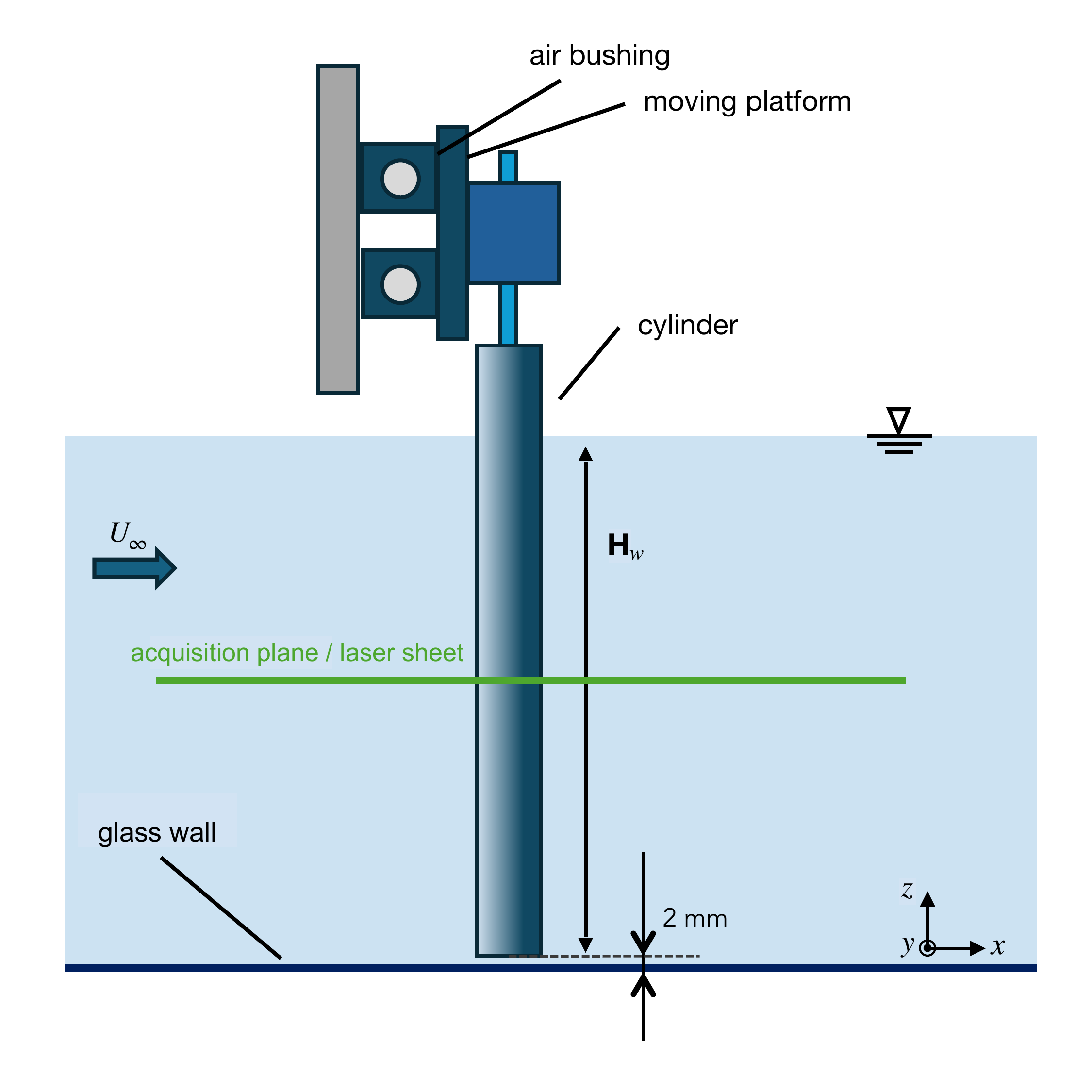}
    \subcaption{}
    \label{fig:02_dataset/exp-set-up}
\end{minipage}
\hspace{2em}
\begin{minipage}[b]{0.45\textwidth}
    \centering
    \includegraphics[width=\linewidth, trim=0cm 0.5cm 0cm 1cm, clip]{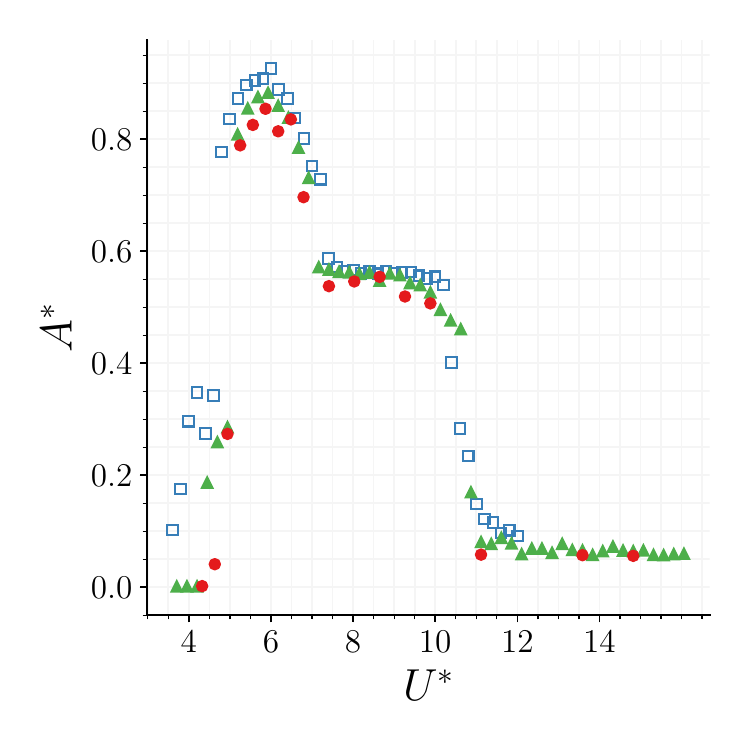}
    \subcaption{}
    \label{fig:02_dataset/amplitude_response}
\end{minipage}
\caption{Experimental setup and cylinder amplitude response. (a) Side-view of the experimental test section setup for Particle Image Velocimetry (PIV) measurements, showing the laser sheet position. (b) Characteristic amplitude response $A^* = A/D$ of the cylinder as a function of reduced velocity $U^* = U_\infty/f_nD$. The red dots indicate the 17 cases acquired in the present study, while results from Schmider et al. \cite{schmider2024improved} (green triangles) and Soti et al. \cite{soti2017harnessing} (blue squares) are included for comparison.}
\end{figure}

The experiments were conducted in the closed-circuit water channel Hydra III at Institut Pprime (Poitiers, France). Figure \ref{fig:02_dataset/exp-set-up} presents a schematic illustration of the experimental setup. The test section measures $2.1\,\text{m}$ in length, $0.51\,\text{m}$ in width, and $0.51\,\text{m}$ in height. The cylinder, which has a length of $0.4\,\text{m}$ and a circular cross-section of diameter $D = 0.05\,\text{m}$, is mounted on an elastically supported platform with negligible damping, allowing it to oscillate freely in the cross-flow direction. The maximum allowed displacement of the cylinder is $\pm 2D$. 

The PIV measurement plane is positioned at mid-height ($H_w/2$) of the immersed portion of the cylinder and covers a region of $4D \times 9D$ in the crosswise ($y$) and streamwise ($x$) directions, respectively. The velocity fields were acquired at a sampling frequency $f_s = 10\,\text{Hz}$, which is an order of magnitude larger than the maximum dominant shedding frequency observed in the data, thereby guaranteeing time-resolved flow structures. The grid spatial resolution is $\Delta x = \Delta y = 0.025D$, ensuring spatially resolved flow measurements.

A total of $N_{\mathrm{oc}}=17$ operating conditions were measured. Figure \ref{fig:02_dataset/amplitude_response} depicts these conditions on the characteristic amplitude response plot of the cylinder, defined by the normalized amplitude $A^* = A/D$ as a function of reduced velocity $U^* = U_\infty/f_nD$, with $U_\infty$ the inflow velocity and $f_n = 0.421\,\text{Hz}$ the cylinder natural oscillation frequency. This response is typical of a low-mass ratio VIV system, and the acquired operating conditions cover all the distinct FSI regimes of interest to VIV energy harvesting devices \cite{govardhan2000modes, schmider2024improved, soti2017harnessing}. Further details about these regimes are reported in Section \ref{sec:02_dataset/2d_flow_snapshots}. A detailed description of the instrumentation and VIV device can be found in Schmider et al. \cite{schmider2024improved}.
\vspace{1cm}

\subsection{Preparation of 2D Flow Snapshots for Model Input}
\label{sec:02_dataset/2d_flow_snapshots}
The acquired PIV measurements define a dataset $\mathcal{D}$ of two-dimensional ($x$-$y$) snapshots including the streamwise and crosswise velocity fields ($u$-$v$). A total of $N_t = 1,000$ snapshots were extracted for each of the $N_{\mathrm{oc}} = 17$ operating conditions shown in Figure \ref{fig:02_dataset/amplitude_response}. The dataset can therefore be written as:

\begin{equation}
    \mathcal{D} = \bigcup_{i=1}^{N_{\mathrm{oc}}} \mathcal{D}_i \, , \quad \text{where} \quad 
    \mathcal{D}_i = \left\{\mathbf{S}_j^{(i)}\right\}_{j=1}^{N_t}
\end{equation}

\noindent with the snapshot $\mathbf{S}_j^{(i)}$ being a structured tensor of shape $N_x \times N_y \times N_c = 416 \times 194 \times 2$. As an extension of this notation, the superscripts Train, Val and Test will be used to denote the data corresponding to the training, validation and test sets defined later in this section. 

The raw PIV data often contain spurious missing values and exhibit long tails in their probability density function, issues that can compromise the network training. To mitigate these, a two-step preprocessing procedure was applied to the PIV acquisitions. For each operating condition $i$, the dataset $\mathcal{D}_i$ is first filtered with Singular Value Decomposition (SVD). Specifically, the singular values were truncated to retain $99$\% of the relative information content, quantified as the ratio of the sum of the retained singular values to the total sum. Second, to remove extreme outliers the probability density functions of each velocity component $c$ in $\mathcal{D}_i$ were clipped at $\mu_c^{(i)} \pm 3\sigma_c^{(i)}$.  Here, $\mu_c^{(i)}$ and $\sigma_c^{(i)}$ are the ensemble mean and standard deviation of the channel $c$ computed as:

\begin{equation}
    \mu_c^{(i)} = \frac{1}{N_t N_x N_y} \sum_{j,k,p} S_{j,k,p,c}^{(i)} \quad, \quad \sigma_c^{(i)} = \sqrt{\frac{1}{N_t N_x N_y} \sum_{j,k,p} \left(S_{j,k,p,c}^{(i)} - \mu_c^{(i)}\right)^2}\quad.
    \label{eq:02_dataset/mean_channel_case}
\end{equation}

\noindent This clipping threshold retains 99.73\% of the original probability density function, assuming a Gaussian distribution.

\begin{figure}[h]
\centering
\begin{minipage}[b]{0.4\textwidth}
    \centering
    \includegraphics[width=\linewidth, trim=0cm 0.5cm 0cm 1cm, clip]{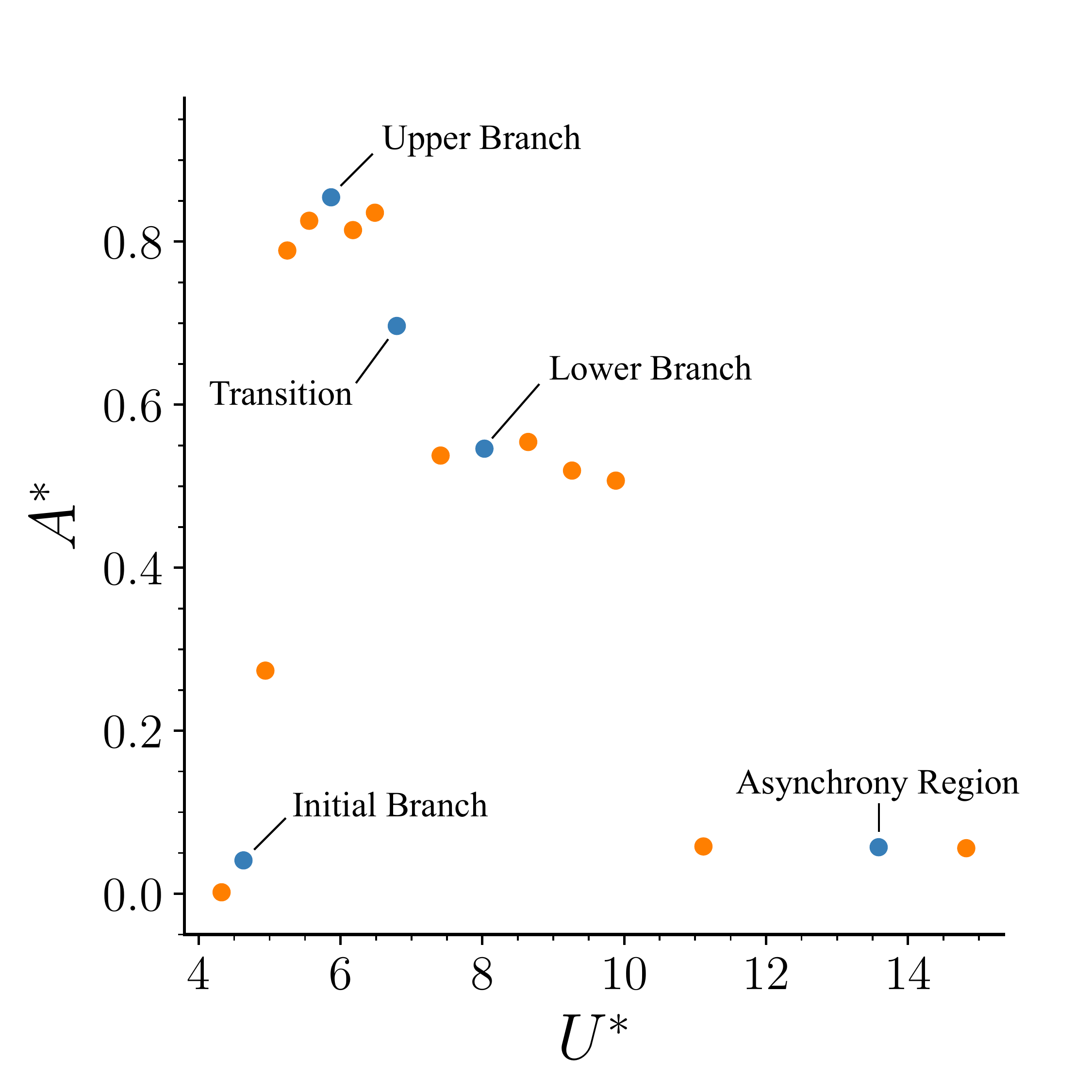}
    \subcaption{}
    \label{fig:02_dataset/dataset_split}
\end{minipage}
\hspace{2em}
\begin{minipage}[b]{0.5\textwidth}
    \centering
    \includegraphics[width=0.48\linewidth, trim=0cm 6cm 0cm 8cm, clip]{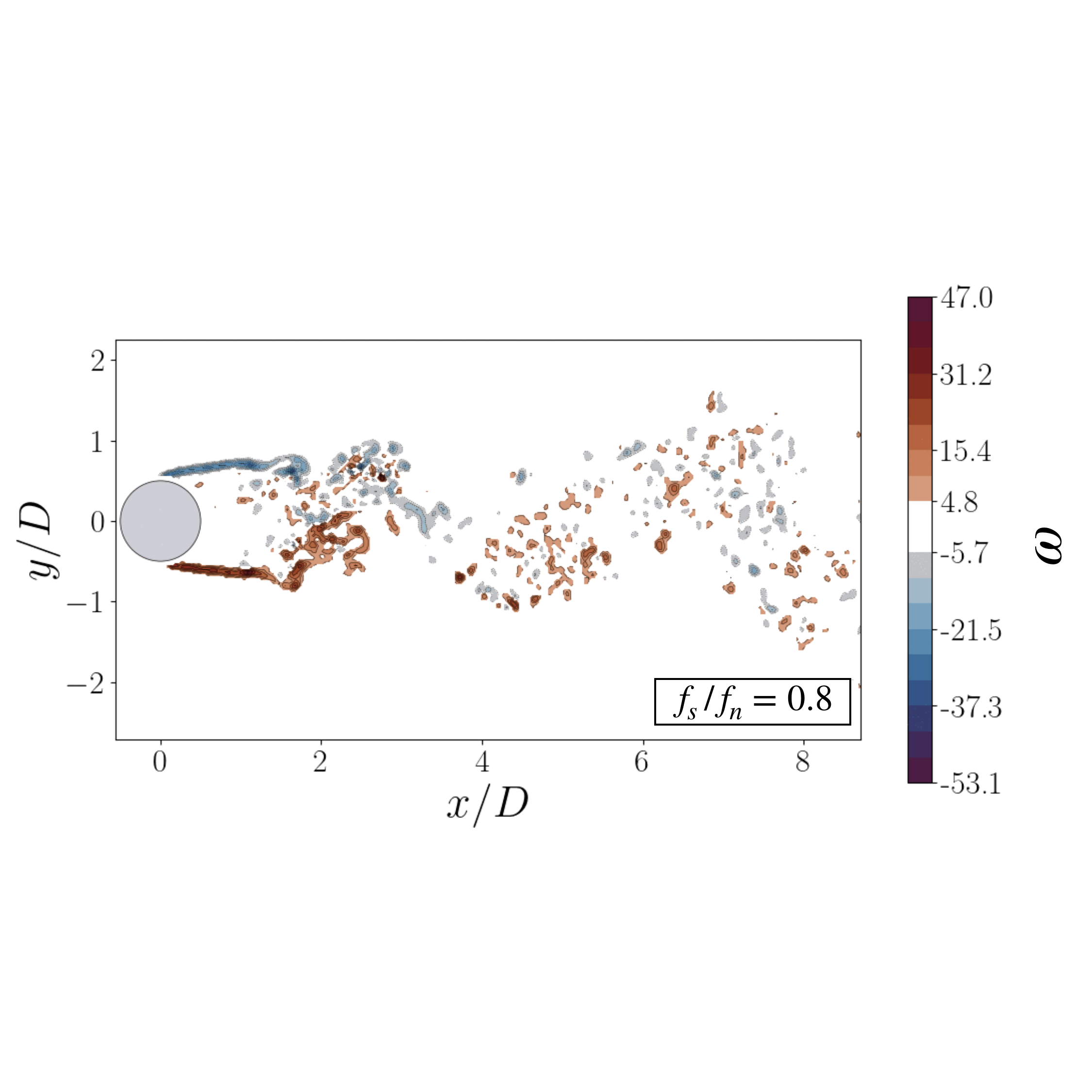}
    \includegraphics[width=0.48\linewidth, trim=0cm 6cm 0cm 8cm, clip]{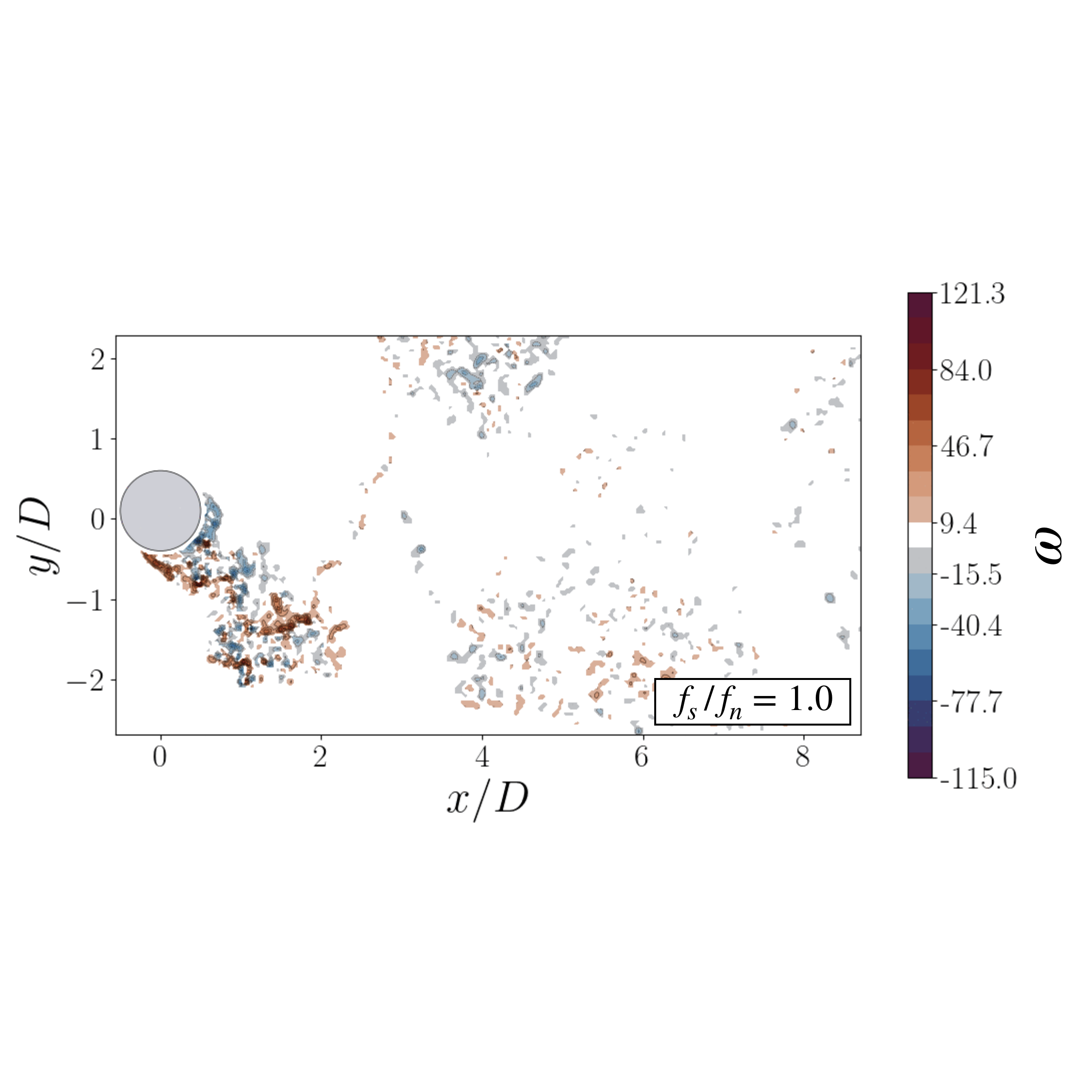}
    \includegraphics[width=0.48\linewidth, trim=0cm 6cm 0cm 8cm, clip]{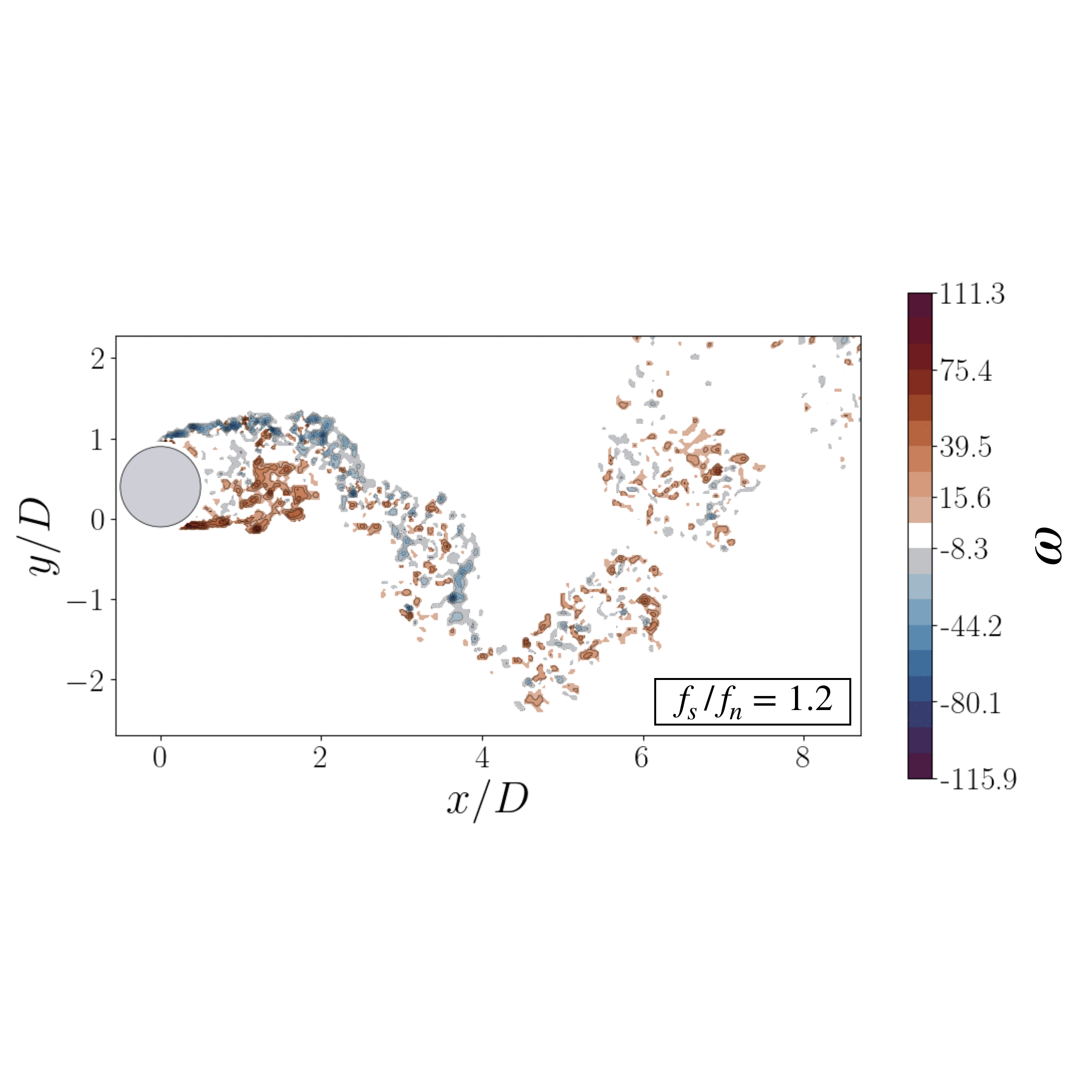}
    \includegraphics[width=0.48\linewidth, trim=0cm 6cm 0cm 8cm, clip]{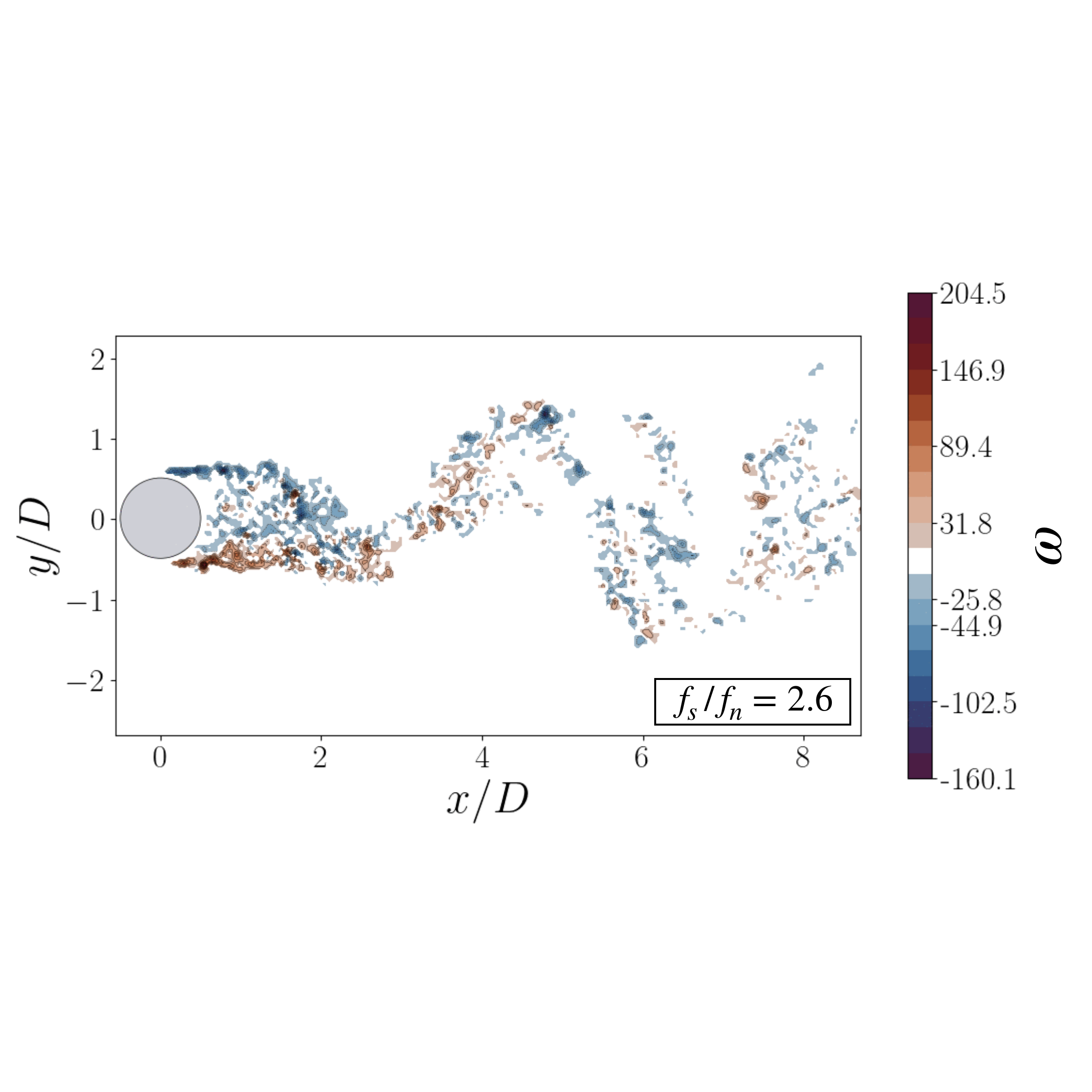}
    \vspace{2em}
    \subcaption{}
    \label{fig:02_dataset/vorticity_cases}
\end{minipage}
\caption{Dataset partitioning strategy and illustrative VIV flow patterns. (a) Amplitude response plot ($A^* = A/D$) versus reduced velocity ($U^* = U_\infty/f_n D$), delineating cases used for training/validation (orange dots) and the test set cases (blue dots). (b) Vorticity ($\omega = \partial v / \partial x - \partial u / \partial y$) visualizations illustrating the diverse fluid-structure interaction regimes within the test set. The normalized vortex shedding frequency ($f_s/f_n$) is also reported for each case. From top-left: Initial, Upper, Lower, and Asynchrony Branches.}
\end{figure}

After this two-step preprocessing the dataset $\mathcal{D}$, containing a total of $17,000$ snapshots, was partitioned into three subsets: a training set $\mathcal{D}^{\mathrm{Train}}$, a validation set $\mathcal{D}^{\mathrm{Val}}$, and a test set $\mathcal{D}^{\mathrm{Test}}$. The training set comprises the first $N_t^{\mathrm{Train}} = 900$ snapshots from the $N_{\mathrm{oc}}^{\mathrm{Train/Val}} = 12$ different operating conditions represented in orange in Figure \ref{fig:02_dataset/dataset_split} (10,800 snapshots in total). The subsequent $N_t^{\mathrm{Val}} = 100$ snapshots from these same operating conditions form the validation set (a total of 1200 snapshots), employed for hyperparameter selection, monitoring training progress against overfitting, and implementing early stopping criteria. Finally, the test set consists of all $N_t^{\mathrm{Test}} = 1,000$ snapshots from the $N_{\mathrm{oc}}^{\mathrm{Test}} = 5$ different operating conditions represented in blue in Figure \ref{fig:02_dataset/dataset_split} (a total of $5,000$ snapshots), strictly reserved for evaluating the performance of the final trained models (results presented in Section~\ref{sec:04_results}).

This test set was designed to be entirely separate from the training and validation sets (orange cases) to rigorously evaluate the model's generalization capability to unseen flow dynamics. Additionally, the test cases were specifically selected to encompass the characteristic fluid-structure interaction regimes present in the dataset:

\begin{itemize}
    \item Initial Branch: This regime occurs at lower reduced velocities, where the vortex shedding frequency is below the structure's natural frequency. The cylinder remains nearly stationary, despite sensing fluid forces.
    \item Upper Branch: As the reduced velocity increases, the vortex shedding frequency begins to match the cylinder's natural frequency, thereby inducing a resonance phenomenon that results in the maximum recorded cylinder oscillation amplitude.
    \item Transition Branch: Further increasing the reduced velocity leads to a shedding frequency slightly higher than the structure's natural frequency. An intermittent regime characterized by switching behavior between the Upper and Lower Branches is therefore observed.
    \item Lower Branch: As reduced velocity (and thus shedding frequency) increases, the cylinder's oscillations synchronize with the shedding frequency (lock-in), following the flow instability.
    \item Asynchrony Region: In this region, the shedding frequency significantly exceeds the natural frequency, leading to a weak coupling of the fluid and structural dynamics. The cylinder displays minimal oscillations.
\end{itemize}

\noindent Figure \ref{fig:02_dataset/vorticity_cases} helps visualize how the flow patterns differ across these distinct branches. It is important to note that the Transition regime is exclusively present in the test set, a higher generalization challenge is thus anticipated for this specific regime. Nevertheless, given that the primary flow mechanisms in this branch represent a combination of those found in the Upper and Lower Branches, the model is still expected to demonstrate generalization capabilities.
% ------------------------ Methods

\section{VIVALDy: A Machine-Learning Framework for Data-Driven Flow Modeling}
\label{sec:03_methods}
This section details the proposed VIVALDy machine-learning framework, describing its components and architectures. The framework operates without explicit knowledge of the operating conditions. Instead, it implicitly infers these from the characteristic features within the input data.

\subsection{Masked Convolutions for Solid-Fluid Interface Fidelity}
A primary challenge in applying convolutional neural networks (CNNs) to fluid domains is handling the presence of structural elements within the convolution field of view, as missing values cannot be backpropagated during training \cite{lecun2015deep}. Encoding and reconstructing these interface regions is essential as this is where flow instabilities are generated and fluid-structure interaction occurs \cite{pope2001turbulent, huera2025vortex}. Excluding such regions would result in loss of relevant flow information during encoding, leading to unsatisfactory reconstructions.

To illustrate, considering the case under study, the presence of the oscillating cylinder must be addressed. A naive approach would be masking the cylinder's grid points with zeros. However, this approach would create ambiguity for the network: a zero value could represent either an actual physical zero (e.g., zero velocity at a stagnation point) or a missing data point within the solid body. Since the cylinder's position changes dynamically across snapshots, this ambiguity might force the network to learn non-physical interpretations. Additionally, since the PIV data exclude the boundary layer, zero masking introduces sharp discontinuities that cause high gradients during backpropagation, leading to unstable training or incorrect inferences. Therefore, this work employs masked CNNs \cite{uhrig2017sparsity}, originally developed for sparse depth estimation, which utilize specialized convolutions with binary masks to distinguish between fluid and solid regions. This approach enables the network to explicitly account for the cylinder's presence and focus calculations solely on the fluid domain.

Formally, let $\mathbf{o}=\{o_{i,j}\}$ be the output of the previous layer and $\mathbf{m}=\{m_{i,j}\}$ the binary mask ($m_{i,j}=1$ if the point of coordinate $(i,j)$ belongs to the fluid and $m_{i,j}=0$ if the point belongs to the solid). The masked convolution operation is defined as:

\begin{equation}
f_{p,q}(\mathbf{o}, \mathbf{m}) \;=\; \frac{\sum\limits_{i,j=-k}^{k} m_{p+i,q+j}\,o_{p+i,q+j}\,w_{ij}}
{\sum\limits_{i,j=-k}^{k}m_{p+i,q+j} + \varepsilon} \;+\; b,
\label{eq:methods/masked_convolution}
\end{equation}

\noindent where $w_{ij}$ are the convolution weights, $k$ is the kernel size, $b$ is a bias term, and $\varepsilon$ is a small constant used to avoid division by zero. When all points under the convolution kernel belong to the fluid (i.e.\ all $m_{p+i,q+j}$ are non-zero), this equation reduces to a standard convolution. The mask $\mathbf{m}$ itself is updated through subsequent layers by max pooling, which allows the network to maintain a consistent notion of valid (fluid) and invalid (solid) regions as the data passes through the network:

\begin{equation}
f_{p,q}^{\mathbf{m}}(\mathbf{m}) \;=\; \max_{\,i,j = -k, \ldots, k}\;m_{p+i,q+j},
\end{equation}

\noindent where $k$ is also the kernel size for the max pooling operation.

\subsection{Latent-Feature Extraction using a $\beta$-VAE-GAN Architecture}

The core architecture of VIVALDy is now presented, comprising a latent-feature extractor designed to extract informative low-dimensional latent features $\boldsymbol{\zeta}$ from high-dimensional flow snapshots $\mathbf{S}$. This architecture incorporates masked convolutions, which effectively manage the solid-fluid interface. The vector $\boldsymbol{\zeta}$ is defined to have a size of $m$, which is a predefined parameter. In this study, the model compresses each $416 \times 194 \times 2$ flow snapshot, defined in Section~\ref{sec:02_dataset/2d_flow_snapshots}, into a latent space of just $m=3$ dimensions, representing an extreme compression ratio exceeding $50,000:1$. This dimensionality was explicitly chosen to enable direct visualization of the latent dynamics without the need for secondary dimensionality reduction (e.g., t-SNE), while maintaining a compact state-space to facilitate the subsequent learning of dynamics.

This extreme compression is achieved while preserving flow statistical properties using a hybrid generative architecture. The proposed design merges $\beta$-Variational Autoencoders \cite{higgins2017beta} with Generative Adversarial Networks \cite{goodfellow2014generative}, motivated by recent works in learned image compression \cite{tschannen2018deep, agustsson2019generative, mentzer2020high}. These works showed how adding adversarial GAN loss to autoencoder-based compression models improves input data distribution preservation in reconstructions. These improvements are obtained through simultaneous optimization of three objectives: reconstruction fidelity, disentanglement of latent variables, and statistical consistency between reconstructed and reference distributions, as will be explained. The following briefly summarizes the foundational components of the proposed architecture.

VAEs are a type of autoencoder that learns \textit{probabilistic} latent-space representations of data. The $\beta$-VAE modification introduces a scaling factor $\beta$ to the Kullback–Leibler (KL) divergence term in the objective function. Assuming the prior distribution $p_{\boldsymbol{\zeta}}$ follows a standard multivariate normal distribution, the encoder $E$ approximates the posterior $p_{\boldsymbol{\zeta}\vert\mathbf{S}}$ via a probability $q_{_{\mathbf{W}_E}}(\boldsymbol{\zeta}\vert\mathbf{S})$, parametrized by the encoder weights $\mathbf{W}_E$. The decoder $G$ then acts as a generative model, approximating $p_{\mathbf{S}\vert\boldsymbol{\zeta}}$ through a distribution $p_{_{\mathbf{W}_G}}(\mathbf{S}\vert\boldsymbol{\zeta})$ parametrized by its weights $\mathbf{W}_G$. The optimization process minimizes a modified evidence lower bound (ELBO),  augmented by $\beta$:

\begin{equation}
    \mathcal{L}_{EG}^{\beta-\mathrm{VAE}} = \mathbb{E}_{\mathbf{S} \sim p_{\mathbf{S}}}\bigl[d(\mathbf{S}, \tilde{\mathbf{S}})\bigr] + \beta\,D_{\mathrm{KL}}\bigl(q_{\mathbf{W}_E}(\boldsymbol{\zeta}\vert\mathbf{S}) \,\|\, p_{\boldsymbol{\zeta}}\bigr),
\label{eq:03_methods/loss_function_beta_VAE}
\end{equation}

\noindent where where $\mathbb{E}[\cdot]$ denotes expectation, $d(\cdot, \cdot)$ is a distortion metric, e.g. the mean square error, and  $D_{\mathrm{KL}}(\cdot)$ is the Kullback--Leibler divergence.

The $\beta$ parameter regulates the trade-off between reconstruction fidelity, $d(\cdot, \cdot)$, and the disentanglement constraint, $D_{\mathrm{KL}}(\cdot)$. Disentanglement refers to the ability to separate independent factors of variation in the data, allowing each latent dimension to capture distinct underlying features. In the context of physical systems, this property implies that variations along specific latent axes correspond to variations in distinct physical parameters \cite{jacobsen2022disentangling}. While assigning specific physical meaning to these axes requires post-hoc analysis (e.g., latent traversals or gradient-based sensitivity analysis), this alignment significantly simplifies the interpretation of the governing dynamics. Higher values of $\beta$ promote greater disentanglement; however, this often comes at the cost of reduced reconstruction fidelity.

Generative Adversarial Networks are probabilistic models composed of two networks, a generator $G$ and a discriminator $D$, trained in opposition to one another. The generator $G(\boldsymbol{\zeta})$ maps latent vectors drawn from a standard Gaussian $\boldsymbol{\zeta} \sim p_{\boldsymbol{\zeta}}$ to reconstructions $\tilde{\mathbf{S}}$, while the discriminator $D$ attempts to distinguish between input samples $\mathbf{S} \sim p_{\mathbf{S}}$ and generated samples $\tilde{\mathbf{S}} \sim p_{\tilde{\mathbf{S}}}$. This two-player min-max game is formalized using the following non-saturating loss:

\begin{align}
    \mathcal{L}_{G}^\mathrm{GAN} &= \mathbb{E}_{\boldsymbol{\zeta}_j \sim p_{\boldsymbol{\zeta}}} \Big[-\,\log\bigl(D\bigl(G(\boldsymbol{\zeta})\bigr)\bigr)\Big], \label{eq:methods/gan_loss_gen}\\
    \mathcal{L}_{D}^\mathrm{GAN} &=\mathbb{E}_{\mathbf{S} \sim p_{\mathbf{S}}} \Big[-\,\log\bigl(D(\mathbf{S})\bigr) -\,\log\bigl(1 - D\bigl(\tilde{\mathbf{S}}\bigr)\bigr) \Big]. 
\label{eq:03_methods/gan_loss_disc}
\end{align} 

The generator's loss, $\mathcal{L}_{G}^\mathrm{GAN}$, represents the negative log probability that the discriminator assigns to the generated data $\tilde{\mathbf{S}} = G(\boldsymbol{\zeta})$ being from the real data distribution. Conversely, the discriminator's loss, $\mathcal{L}_{D}^\mathrm{GAN}$, minimizes two terms: the negative log probability that it assigns to real data $\mathbf{S}$ being real (first term), and the negative log probability that it assigns to generated data being fake $\tilde{\mathbf{S}}$ (second term). This adversarial loss formulation, where the generator tries to minimize $\mathcal{L}_{G}^\mathrm{GAN}$ and the discriminator tries to minimize $\mathcal{L}_{D}^\mathrm{GAN}$, drives the generator to produce outputs that are statistically consistent with the real data distribution, $p_{\tilde{\mathbf{S}}} \approx p_\mathbf{S}$.

\begin{figure}
    \centering
    \includegraphics[width=0.9\linewidth, trim=0.5cm 8cm 1.0cm 8cm, clip]{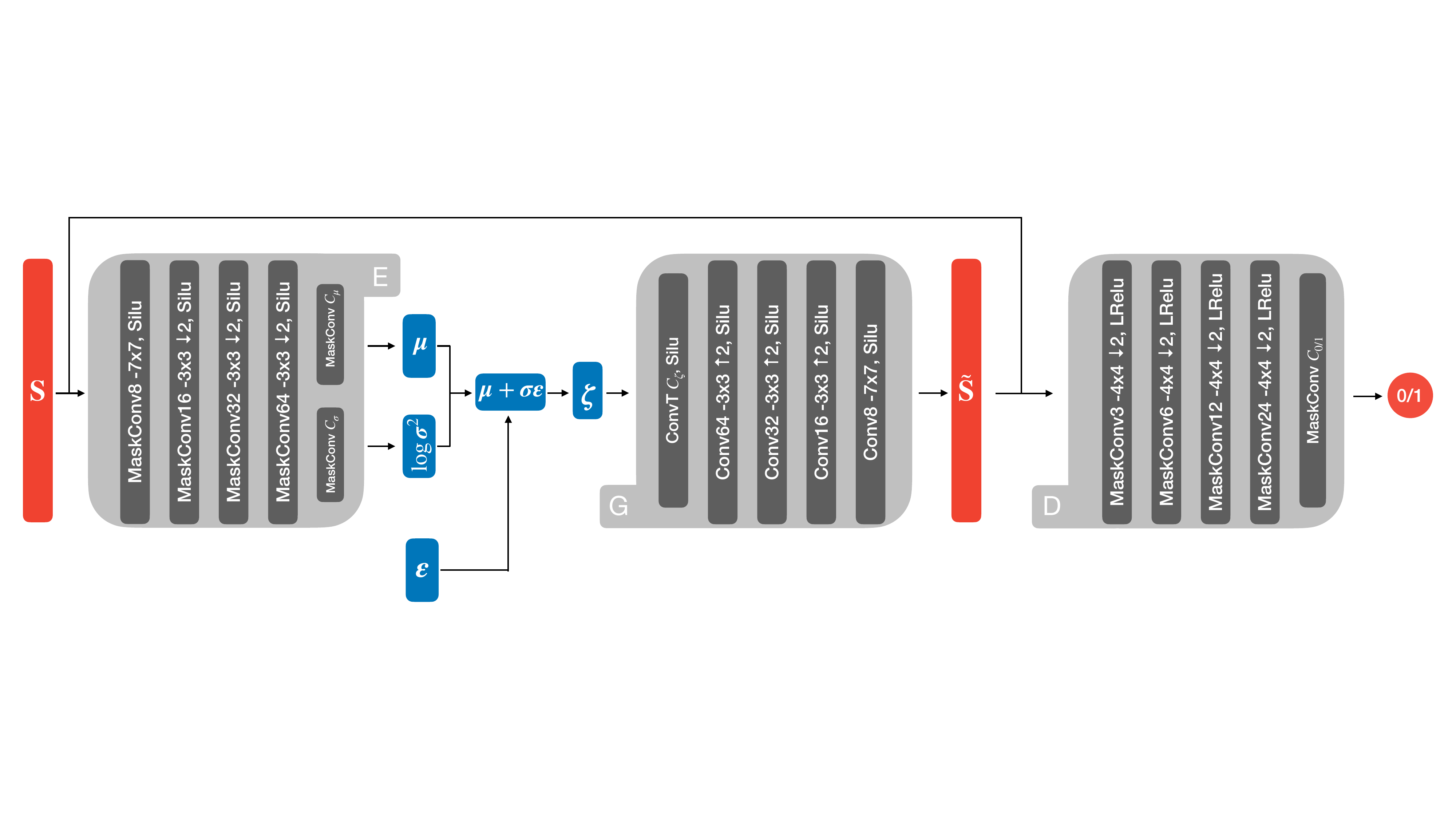}
    \caption{
    $\beta$-VAE-GAN architecture. The encoder ($E$), generator ($G$), and discriminator ($D$) use (masked) convolutional layers. Each (Mask)Conv$C$ layer applies $k \times k$ filters with $C$ output channels, where $k$ is the filter kernel size. Downward arrows (↓2) denote strided down-sampling, while upward arrows (↑2) denote Lanczos up-sampling. Silu (\textit{sigmoid linear unit}) and LRelu (\textit{leaky ReLU}) are used as activation functions. The layers labelled \texttt{Conv~$C_\mu$} and \texttt{Conv~$C_\sigma$} project their inputs to the mean and variance of the latent variables, respectively, whereas \texttt{ConvT~$C_{\zeta}$} projects the sampled latent variable $\boldsymbol{\zeta}$ back to the encoder’s final layer dimensionality.
    }
    \label{fig:03_methods/advae_arch}
\end{figure}

The combined $\beta$-VAE-GAN framework (illustrated in Figure~\ref{fig:03_methods/advae_arch}) consists of three components: an encoder $E$, a decoder/generator $G$, and a discriminator $D$. The \emph{masked convolutions}, introduced in the previous subsection are employed in the encoder and discriminator, to explicitly account for the cylinder presence in the flow domain. In contrast, the decoder does not require masking, as the masked convolutions in the encoder ensure that only the fluid region grid points contribute to its inputs (the latent variables representing the flow region). The encoder and generator are jointly optimized, while being trained adversarially against the discriminator. The training objective integrates terms from both the $\beta$-VAE and the GAN formulations into the following adversarial loss:

\begin{align}
    \mathcal{L}_{EG}^\mathrm{Hybrid}  &= \underbrace{\mathbb{E}_{\mathbf{S} \sim p_{\mathbf{S}}}\bigl[d(\mathbf{S}, \tilde{\mathbf{S}})\bigr] + \beta\,D_{\mathrm{KL}}\bigl(q_{\mathbf{W}_E}(\boldsymbol{\zeta}\vert\mathbf{S}) \,\|\, p_{\boldsymbol{\zeta}}\bigr)}_{\text{$\beta$-VAE terms}} \;-\;\alpha\,\underbrace{\mathbb{E}_{\mathbf{S} \sim p_{\mathbf{S}}}\bigl[\log\bigl(D(\tilde{\mathbf{S}})\bigr)\bigr]}_{\text{GAN term}}, 
    \label{eq:adversarial_loss_formulation/enc_gen_loss}\\ \vspace{5ex}
    \mathcal{L}_{D}^\mathrm{Hybrid} &= \underbrace{\mathbb{E}_{\mathbf{S} \sim p_{\mathbf{S}}} \bigl[-\,\log\bigl(D(\mathbf{S})\bigr) -\,\log\bigl(1 - D\bigl(\tilde{\mathbf{S}}\bigr)\bigr)\bigr]}_{\text{GAN term}}.
    \label{eq:adversarial_loss_formulation/dsc_loss}
\end{align}

Here, $\beta$ and $\alpha$ are user-defined hyperparameters that control the trade-off between reconstruction accuracy $d(\cdot,\cdot)$, latent disentanglement $D_{\mathrm{KL}}(\cdot)$, and statistical fidelity $\log D(\cdot)$. During training, two separate optimizers update the networks parameters: one for the combined $EG$ loss (Equation~\ref{eq:adversarial_loss_formulation/enc_gen_loss}) and another one for the discriminator (D) loss (Equation~\ref{eq:adversarial_loss_formulation/dsc_loss}). Training details are provided in Section \ref{sec03/training_details}.

\subsection{Latent-Space Dynamics from Cylinder Kinematics using a Bidirectional Transformer}

The $\beta$-VAE-GAN architecture provides spatial compression for each snapshot. The physical system dynamics are consequently represented as an evolving trajectory within the low-order latent space. To reconstruct the full spatio-temporal dynamics from cylinder displacement alone, a separate architecture that maps this displacement directly to the target trajectory in the latent space is introduced. When coupled with the $\beta$-VAE-GAN decoder, this architecture enables reconstruction of the original high-dimensional flow fields.

The model's objective is to learn a mapping, $T$, that predicts the flow's trajectory in the low-dimensional latent space from the cylinder's displacement. Given an input sequence of cylinder displacements over a time window of length $H$, denoted as $\mathbf{y}_{\mathrm{cyl}} = [y_{t-H+1}, \dots, y_t]$, the model predicts the corresponding latent space trajectory $\boldsymbol{\zeta} \in \mathbb{R}^{m\times H}$. This output sequence is denoted as $\boldsymbol{\zeta} = [\boldsymbol{\zeta}_{t-H+1}, \dots, \boldsymbol{\zeta}_t]$, where each entry $\boldsymbol{\zeta}_k \in \mathbb{R}^m$ represents the compressed state vector of the flow at time step k. Note that while the input is a sequence of scalars (1-DOF displacement), the output is a sequence of multi-dimensional state vectors (here $m=3$). The complete function is defined as:

\begin{equation}
	\boldsymbol{\zeta} = T(\mathbf{y}_{\mathrm{cyl}}).
\end{equation}

Physically, this mapping $T$ can be interpreted as the learned coupling between the structural motion of the cylinder and the dynamics of the surrounding flow field.

\begin{figure}[h]
    \centering
    \includegraphics[width=0.9\linewidth, trim=0cm 8cm 0cm 8cm, clip]{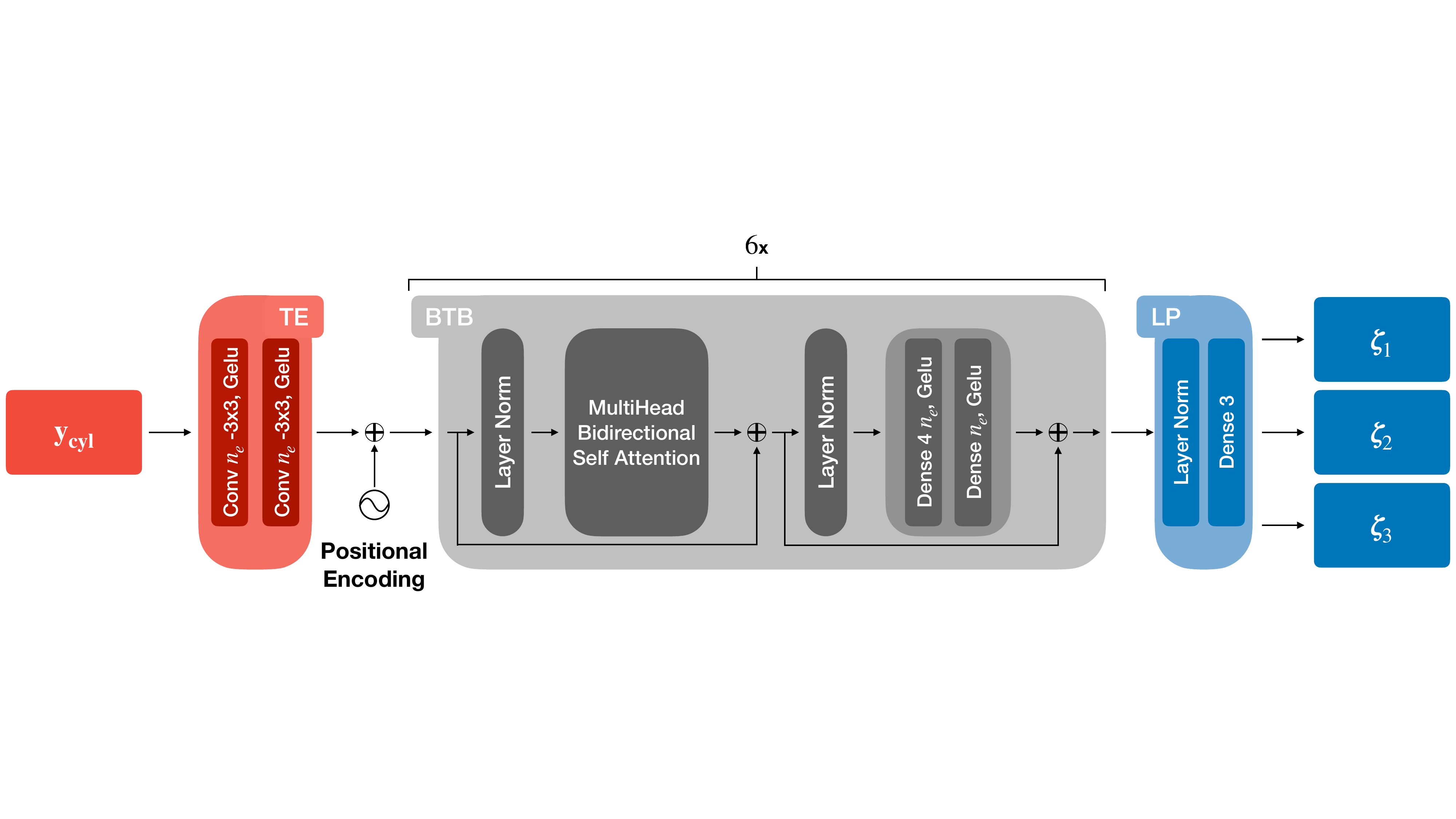}
    \caption{
    Bidirectional transformer architecture. The time embedding (TE) layer consists of two convolutional layers with $3\times3$ filters and Gelu (Gaussian error linear unit) activation functions. The number of filter channels is set to $n_e$, which determines the dimensionality of the encoded representation fed into the bidirectional transformer block (BTB). The latter uses a multihead, bidirectional self-attention mechanism, following the original transformer formulation \cite{vaswani2017attention}, combined with a shallow dense feedforward network. This transformer block is repeated six times. Finally, a dense latent projection (LP) layer maps the transformer's output to the latent trajectory space.
    }
    \label{fig:03_methods/bitransformer}
\end{figure}

The problem is solved using an encoder-only bidirectional transformer (Figure~\ref{fig:03_methods/bitransformer}), drawing inspiration from Bidirectional Encoder Representations from Transformers (BERT) \cite{devlin2019bert}. The architecture comprises three primary components: a time-embedding (TE), a bidirectional transformer block (BTB), and a linear projection (LP). The TE component consists of a two-layer CNN that encodes the input displacement sequence to a higher-dimensional representation of dimension $n_e$. Positional encoding is then added to provide the sequence order information, and the resulting tensor is passed to the BTB. The BTB consists of six transformer encoder layers, each containing six-head self-attention mechanisms combined with feedforward networks. The self-attention mechanism operates bidirectionally without masking, allowing each time step to simultaneously consider information from all other time steps in the sequence, following the original transformer encoder formulation by Vaswani \textit{et al} \cite{vaswani2017attention}. Finally, the output passes through a linear projection layer, consisting of a one-layer perceptron, which maps the BTB output to the desired latent space trajectory dimension. Two hyperparameters control the model configuration: the TE output dimension $n_e$ and a parameter $n_H$ defining the maximum context length. The latter specifies the maximum number of time steps the attention mechanism can process simultaneously. While the model can handle sequences of arbitrary length $H$ through windowing strategies, attention computations are constrained to this maximum context length. Further training and hyperparameter details are reported in Section \ref{sec03/training_details}.

\begin{figure}[h]
    \centering
    \includegraphics[width=\linewidth, trim=0cm 12cm 0cm 12cm, clip]{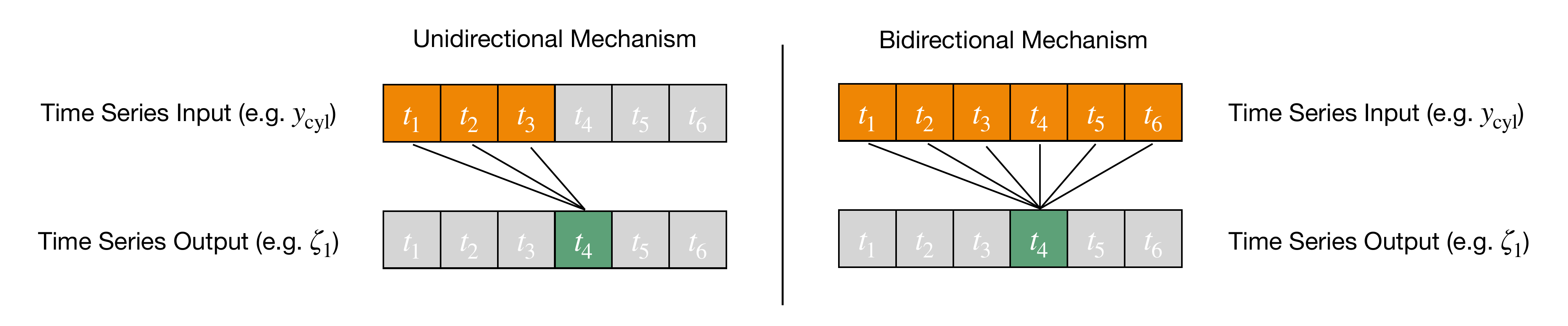}
    \caption{Comparison of unidirectional and bidirectional attention mechanisms for time series. In a unidirectional mechanism (left), the output at time $t_4$ can only attend to inputs before $t_4$ (e.g., $t_j$ where $j < 4$). This masking restricts the model's ability to capture lead-lag relationships. Conversely, a bidirectional mechanism (right) allows the output at $t_i$ to attend to all input timesteps.}
    \label{fig:03_methods/bidirectional_scheme}
\end{figure}

The design choice of using bidirectional over unidirectional attention is better understood when considering the simplified scheme presented in Figure \ref{fig:03_methods/bidirectional_scheme}. The objective is to learn the mapping between output and input time series. When mapping a general time instant $t_i$ (for example $t_4$ in the figure), a unidirectional attention mechanism constrains the output $t_i$ to depend only on input elements prior to its time step ($t_j$ with $j < i$). This constraint becomes problematic when significant temporal lags or leads exist between the input time series (cylinder kinematics) and the output time series (latent flow dynamics), as it prevents the model from accessing the full-context information. In contrast, a bidirectional attention mechanism operates without such masking, allowing the model to utilize information from the complete input time series. Additionally, if the underlying problem exhibits inherent unidirectional dependencies, the bidirectional mechanism can converge to unidirectional behavior during training. By utilizing the full temporal context of the cylinder's displacement signal, this approach enhances the predictive accuracy of latent-space trajectories.

\subsection{Models Training Details}
\label{sec03/training_details}

All hyperparameters used during the training of the $\beta$-VAE-GAN and bidirectional transformer are detailed in Table \ref{tab:03_methods/optimizer_hyperparameters}.

\begin{table}[h!]
    \centering
    \begin{tabular}{l c l c}  
    \toprule\toprule
        \multicolumn{4}{c}{Optimizer Encoder/Generator}\\
        \midrule
        Batch Size ($N_b$) & 32 & Warmup Steps ($ws$) & $100$ \\
        Initial learning rate ($\eta_{\mathrm{start}}$) & $2\times 10^{-2}$ & Initial $\beta$ ($\beta_{\mathrm{start}}$) & $0.0$ \\
        Final learning rate ($\eta_{\mathrm{end}}$) & $2\times 10^{-5}$ & Final $\beta$ ($\beta_{\mathrm{end}}$) & $1\times 10^{-4}$ \\
        Number of cycles ($N_{\mathrm{cycles}}$) & $2$ & Cycle ratio ($R$)& $0.5$\\
        \midrule
        \multicolumn{4}{c}{Optimizer Discriminator}\\
        \midrule
        Batch Size ($N_b$) & 32 & Warmup Steps ($ws$) & $100$ \\
        Initial learning rate ($\eta_{\mathrm{start}}$) & $2\times 10^{-2}$ & Final learning rate ($\eta_{\mathrm{end}}$) & $2\times 10^{-5}$ \\
        \midrule
        \multicolumn{4}{c}{Optimizer Transformer}\\
        \midrule
        Batch Size ($N_b$) & 64 & Warmup Steps ($ws$) & $100$ \\
        Initial learning rate ($\eta_{\mathrm{start}}$) & $3\times 10^{-4}$ & Final learning rate ($\eta_{\mathrm{end}}$) & $3\times 10^{-8}$\\
        Dropout rate ($dr$) & $2\times 10^{-1}$ & \\
    \bottomrule\bottomrule
    \end{tabular}
    \vspace{0.5cm}
    \caption{Optimizer hyperparameters used for training of the $\beta$-VAE-GAN (Figure \ref{fig:03_methods/advae_arch}) and bidirectional-transformer (Figure \ref{fig:03_methods/bitransformer}).}
    \label{tab:03_methods/optimizer_hyperparameters}
\end{table}

The $\beta$-VAE-GAN architecture was trained on $\mathcal{D}^{\mathrm{Train}}$ (Section \ref{sec:02_dataset/2d_flow_snapshots}) using a mini-batch ADAM optimizer with random shuffling. The model was trained for $2,500$ epochs on a single NVIDIA V100 32GB GPU on the Jean-Zay CNRS supercomputer. During training, the learning rate was adjusted using a warmup cosine decay schedule, and the $\beta$ hyperparameter was cyclically annealed \cite{fu2019cyclical}. An early stopping strategy was adopted to prevent overfitting by monitoring the loss on $\mathcal{D}^{\mathrm{Val}}$. The discriminator weights were updated every 5 epochs and frozen during the remaining epochs to stabilize the GAN training dynamic and prevent the discriminator from dominating the generator. The hyperparameter $\alpha$ was kept constant throughout training. An ablation study on this parameter was conducted, training models with $\alpha = \{0.2, 0.02, 0.002\}$ and comparing against an equally trained $\beta$-VAE ($\alpha = 0$), which served as reference, to assess its impact on performance. Values of $\alpha > 0.2$ were excluded from the final analysis as they induce training instabilities in the present setup. Specifically, the adversarial loss was observed to dominate the optimization, causing the generator to produce significant artifacts unrelated to physical flow structures. The results of this $\alpha$ sensitivity study are reported in Section \ref{sec:04_1}.

The trained $\beta$-VAE-GAN encoder (specifically, the $\alpha = 0.2$ model which showed the best performance in the ablation study) was then used to encode $\mathcal{D}^{\mathrm{Train}}$ into the learned low-dimensional space. The obtained target trajectories $\boldsymbol{\zeta}^{\mathrm{gt}}$ were divided into target sequences of size $H = 64$. The same partitioning was applied to the corresponding cylinder displacement acquired during the experiment. The obtained input/output pairs were used to define the training dataset for the bidirectional transformer (Figure \ref{fig:03_methods/bitransformer}). The model context length was set to $n_H = 64$, while the temporal encoding dimension $n_e = 36$. The model was optimized using a mini-batch ADAM optimizer with a warmup cosine decay schedule for the learning rate. An early stopping strategy was adopted to prevent overfitting by monitoring the loss on input/output pairs obtained following the same procedure with $\mathcal{D}^{\mathrm{Val}}$. The training was conducted for  $10,000$ epochs on a single NVIDIA V100 32GB GPU at the Jean-Zay CNRS supercomputer.

% ------------------------------- Results
\section{Results}
\label{sec:04_results}

\subsection{Ablation Study on Adversarial Loss Term}
\label{sec:04_1}

This section reports results from the ablation study on hyperparameter $\alpha$, which controls the weight of the adversarial loss component in Eq. \ref{eq:adversarial_loss_formulation/enc_gen_loss}. This investigation focuses on understanding how adversarial training affects the trade-off between reconstruction fidelity and distributional alignment in the proposed hybrid architecture.

\subsubsection{Evaluation metrics}
\label{sec:04_1/evaluation_metrics}
Two metrics are evaluated on the test set $\mathcal{D}^{\mathrm{Test}}$ to assess model performance. Reconstruction accuracy is measured using Normalized Root Mean Square Error (NRMSE). For each test case $i$ in $\mathcal{D}^{\mathrm{Test}}$ and velocity component $c$, the NRMSE is defined as:

\begin{equation}
    \text{NRMSE}^{(i)}_c = \frac{\sqrt{\frac{1}{N_t N_x N_y} \sum_{j,k,p} \left( S_{j,k,p,c}^{{\prime}^{(i)}} - \tilde{S}_{j,k,p,c}^{{\prime}^{(i)}} \right)^2}}{\max_{j,k,p}(S_{j,k,p,c}^{{\prime}^{(i)}}) - \min_{j,k, p} (S_{j,k,p,c}^{{\prime}^{(i)}})}
\end{equation}

\noindent where $S_{j,k,p,c}^{{\prime}^{(i)}}$ represents fluctuation velocity with mean $\mu^{(i)}_c$ (defined in Eq. \ref{eq:02_dataset/mean_channel_case}) subtracted, and $j,k,p$ are the time, x-direction and y-direction indices, respectively.

Distributional alignment is quantified using the Wasserstein distance, which measures the minimum cost of transforming one distribution into another, making it robust for comparing complex flow field distributions. Smaller values indicate better alignment between generated and true distributions. For each test case $i$ and velocity component $c$, the Wasserstein distance is:

\begin{equation}
    W^{(i)}_c = \inf_{\gamma \in \Pi\left(p(\mathbf{S}^{(i)}_c), \, p(\tilde{\mathbf{S}}^{(i)}_c)\right)} \left(\mathbb{E}_{(\mathbf{S}^{(i)}_c, \tilde{\mathbf{S}}^{(i)}_c) \sim \gamma}\left[ \| \mathbf{S}^{(i)}_c- \tilde{\mathbf{S}}^{(i)}_c \| \right]\right)
\end{equation}

\noindent where $\|\cdot\|$ is the Euclidean norm, and $p(\mathbf{S}^{(i)}_c)$, $p(\tilde{\mathbf{S}}^{(i)}_c)$ are the empirical distributions of true and reconstructed velocity fields, respectively. The infimum is taken over all joint distributions $\gamma \in \Pi(p(\mathbf{S}^{(i)}_c), p(\tilde{\mathbf{S}}^{(i)}_c))$ with the specified marginals.

\subsubsection{Results analysis}

Table \ref{tab:04_1/results_ablation} reports the ablation study results for both velocity components, revealing varied impacts of $\alpha$. Percentages indicate changes relative to the $\alpha = 0$ model, where negative and positive values represent improvements and degradation, respectively. For brevity, the $i$ index denoting test cases is dropped, with $\text{NRMSE}_u$ and $W_u$ denoting respective metrics for the $u$-component, and similarly for the $v$-component.

For $\text{NRMSE}_u$, the $\alpha=0.2$ configuration shows a 8.64\% improvement in the transition regime and smaller gains in initial (0.48\%) and lower (0.60\%) regimes. Conversely, the other $\alpha$ values produce minimal changes, with $\alpha=0.002$ showing small degradations in some regimes (e.g., 1.29\% increase for the initial). $\text{NRMSE}_v$ follows similar trends, with $\alpha=0.2$ achieving the most substantial improvement of 15.90\% in the transition regime and moderate gains in upper (1.60\%) and initial (0.75\%) regimes. Lower $\alpha$ values again yield less favorable results. The $\alpha=0.2$ configuration emerges as most effective for reconstruction accuracy, particularly excelling in the Transition regime.

\begin{table}[h!]
    \centering
    \begin{tabular}{llcccc}
    \toprule\toprule
        Metric & Case & $\alpha$=0.0 & $\alpha$=0.2 & $\alpha$=0.02 & $\alpha$=0.002 \\
        \midrule
        \multirow{5}{*}{$\text{NRMSE}_u$}
        & Initial    & 0.0621 & 0.0618 (-0.48\%) & 0.0629 (+1.29\%) & 0.0627 (+0.97\%) \\
        & Upper      & 0.0927 & 0.0926 (-0.11\%) & 0.0929 (+0.22\%) & 0.0925 (-0.22\%) \\
        & Transition & 0.1088 & \textcolor{green}{0.0993 (-8.64\%)} & 0.1090 (+0.18\%) & 0.1087 (-0.09\%) \\
        & Lower      & 0.0842 & 0.0837 (-0.60\%) & 0.0843 (+0.12\%) & 0.0843 (+0.12\%) \\
        & Asynchrony & 0.0752 & 0.0752 (+0.00\%) & 0.0760 (+1.06\%) & 0.0758 (+0.80\%) \\
        \midrule
        \multirow{5}{*}{$\text{NRMSE}_v$}
        & Initial    & 0.0935 & 0.0928 (-0.75\%) & 0.0954 (+2.03\%) & 0.0961 (+2.78\%) \\
        & Upper      & 0.1061 & 0.1044 (-1.60\%) & 0.1058 (-0.28\%) & 0.1053 (-0.75\%) \\
        & Transition & 0.1151 & \textcolor{green}{0.0968 (-15.90\%)} & 0.1170 (+1.65\%) & 0.1158 (+0.61\%) \\
        & Lower      & 0.0898 & 0.0880 (-2.00\%) & 0.0910 (+1.34\%) & 0.0900 (+0.22\%) \\
        & Asynchrony & 0.0988 & 0.0986 (-0.20\%) & 0.0996 (+0.81\%) & 0.1003 (+1.52\%) \\
        \midrule
        \multirow{5}{*}{$W_u$}
        & Initial    & 0.0027 & 0.0029 (+7.41\%) & 0.0028 (+3.70\%) & \textcolor{red}{0.0031 (+14.81\%)} \\
        & Upper      & 0.0036 & \textcolor{red}{0.0044 (+22.22\%)} & 0.0038 (+5.56\%) & \textcolor{red}{0.0046 (+27.78\%)} \\
        & Transition & 0.0058 & 0.0060 (+3.45\%) & 0.0058 (+0.00\%) & \textcolor{red}{0.0068 (+17.24\%)} \\
        & Lower      & 0.0045 & 0.0047 (+4.44\%) & 0.0041 (-8.89\%) & 0.0049 (+8.89\%) \\
        & Asynchrony & 0.0087 & 0.0087 (+0.00\%) & 0.0080 (-8.05\%) & 0.0094 (+8.05\%) \\
        \midrule
        \multirow{5}{*}{$W_v$}
        & Initial    & 0.0026 & \textcolor{green}{0.0022 (-15.38\%)} & \textcolor{green}{0.0019 (-26.92\%)} & \textcolor{green}{0.0023 (-11.54\%)} \\
        & Upper      & 0.0056 & \textcolor{green}{0.0048 (-14.29\%)} & \textcolor{green}{0.0042 (-25.00\%)} & 0.0051 (-8.93\%) \\
        & Transition & 0.0072 & \textcolor{green}{0.0049 (-31.94\%)} & \textcolor{green}{0.0060 (-16.67\%)} & 0.0069 (-4.17\%) \\
        & Lower      & 0.0063 & \textcolor{green}{0.0048 (-23.81\%)} & \textcolor{green}{0.0048 (-23.81\%)} & 0.0061 (-3.17\%) \\
        & Asynchrony & 0.0087 & \textcolor{green}{0.0078 (-10.34\%)} & \textcolor{green}{0.0065 (-25.29\%)} & 0.0083 (-4.60\%) \\
    \bottomrule\bottomrule
    \end{tabular}
    \vspace{0.5cm}
    \caption{Results of the ablation study on the hyperparameter $\alpha$, controlling the weight of the adversarial loss component (Eq. \ref{eq:adversarial_loss_formulation/enc_gen_loss}).}
    \label{tab:04_1/results_ablation}
\end{table}

Wasserstein distance results reveal asymmetric performance between velocity components. For $W_u$, $\alpha=0.2$ unexpectedly degrades performance in the upper branch (22.22\% increase) while showing minor increases for the other regimes. In contrast, $\alpha=0.02$ provides improvements in lower (8.89\% reduction) and asynchrony (8.05\% reduction) regimes. The $\alpha=0.002$ configuration consistently increases $W_u$ values, indicating minimal adversarial weighting offers little benefit. $W_v$ results present a more favorable picture for $\beta$-VAE-GAN models. All $\alpha$ configurations show improvements across most cases, with $\alpha=0.02$ achieving the largest gains: 26.92\% in initial, 25.00\% in upper, and 25.29\% in asynchrony. The $\alpha=0.2$ model also delivers strong improvements, including a substantial 31.94\% reduction in the transition case. Even $\alpha=0.002$ provides consistent but modest improvements.

Overall, the adversarial term seems to improve the distributional alignment of the cross-stream velocity component $v$, sometimes at the expense of worse $u$ distribution alignment. The $\alpha=0.2$ configuration provides the optimal balance, delivering better reconstruction accuracy while maintaining reasonable distributional performance. 

\subsection{Latent-Space Trajectory Prediction}
\label{sec:04_2}
Having established $\alpha = 0.2$ as the optimal adversarial loss configuration in the preceding analysis, this section examines the bidirectional transformer's capability to map cylinder displacement $\mathbf{y}_{\mathrm{cyl}}$ to target latent-space trajectories $\boldsymbol{\zeta}$ encoded by the $\beta$-VAE-GAN.

Two cases from the test subset (Section \ref{sec:02_dataset/2d_flow_snapshots}) are analyzed in detail: the upper branch ($U^* = 5.56$) and the lower branch ($U^* = 8.03$). These operating conditions are typical of the extended synchronization region characterizing low-mass VIV systems. They were selected because, although both cases lie within the resonance region, they exhibit different dynamics. Specifically, the lower branch case is characterized by highly periodic oscillations, while the upper branch exhibits an irregular chaotic dynamics \cite{khalak1999motions, zhao2014chaotic}.

\subsubsection{Methodological Framework}
All analyses use normalized representations to facilitate quantitative comparison. The predicted and target latent variables are normalized by the maximum magnitude of the target latent variables according to:

\begin{equation}
	\zeta_{j,i}^{\mathrm{norm}} = \zeta_{j,i} / \max_j \left(\sqrt{(\zeta_{j,1}^\mathrm{gt})^2 + (\zeta_{j,2}^\mathrm{gt})^2 + (\zeta_{j,3}^\mathrm{gt})^2}\right)
\end{equation}

\noindent where $\zeta_{j,i}$ is the $i$-th component of the latent vector at time step $t_j$, and $\zeta_{j,i}^\mathrm{gt}$ represents the target trajectory obtained from $\beta$-VAE-GAN encoding of PIV flow field snapshots.

\subsubsection{Temporal Evolution and Statistical Properties}
\label{sec:04_2/Temporal_Evolution}

\begin{figure}[h!]
    \centering
    \begin{subfigure}[b]{0.47\textwidth}
        \centering
        \includegraphics[width=\textwidth]{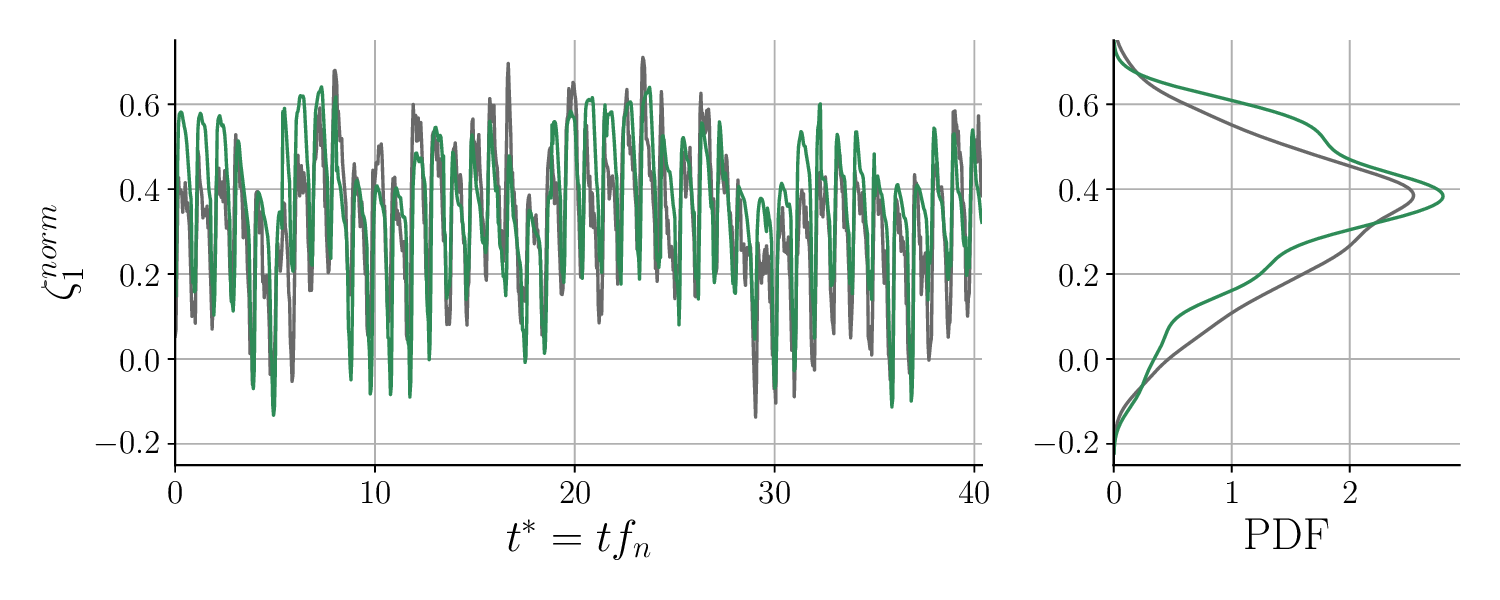}
        \includegraphics[width=\textwidth]{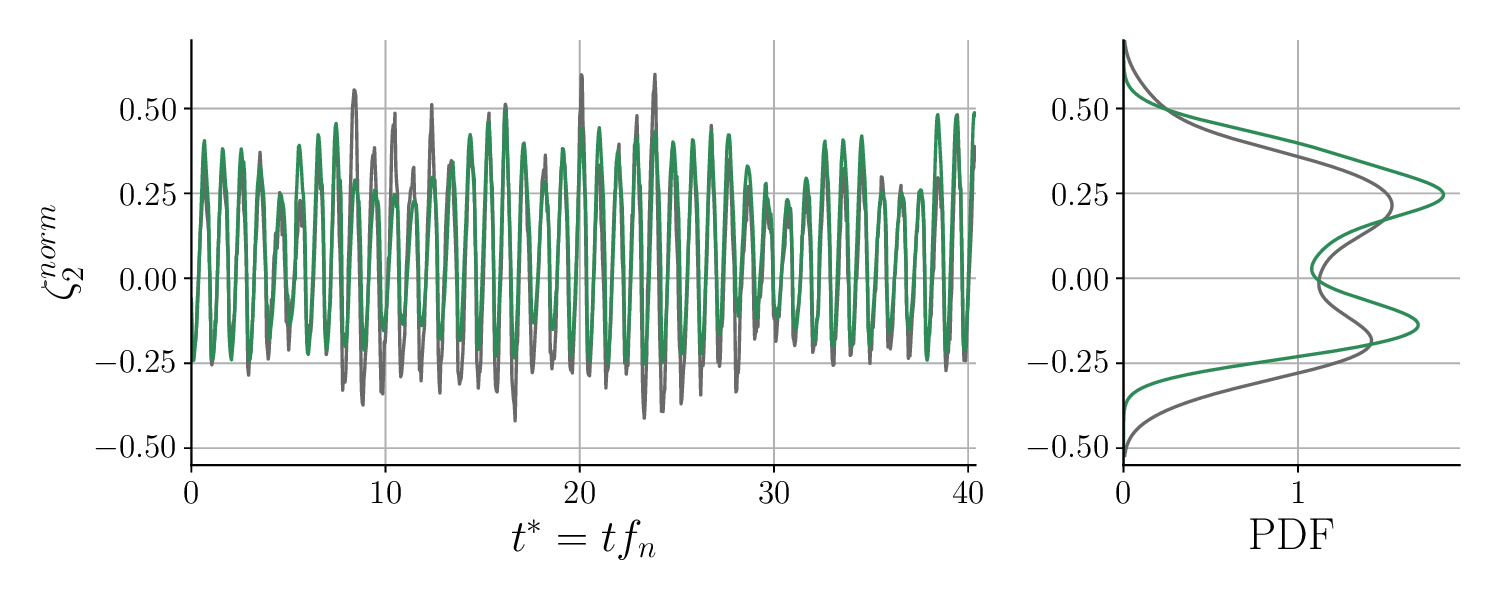}
        \includegraphics[width=\textwidth]{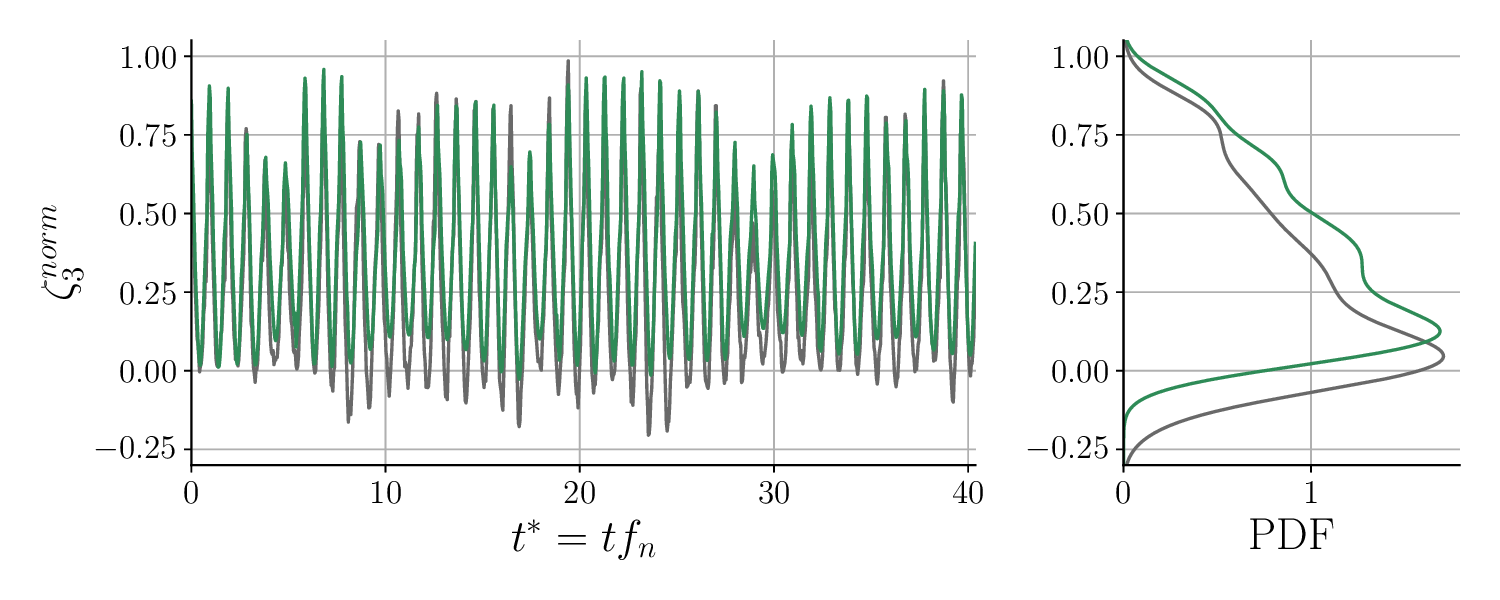}
        \subcaption{}
        \label{fig:04.2/time_series_0556}
    \end{subfigure}
    \hfill
    \begin{subfigure}[b]{0.47\textwidth}
        \centering
        \includegraphics[width=\textwidth]{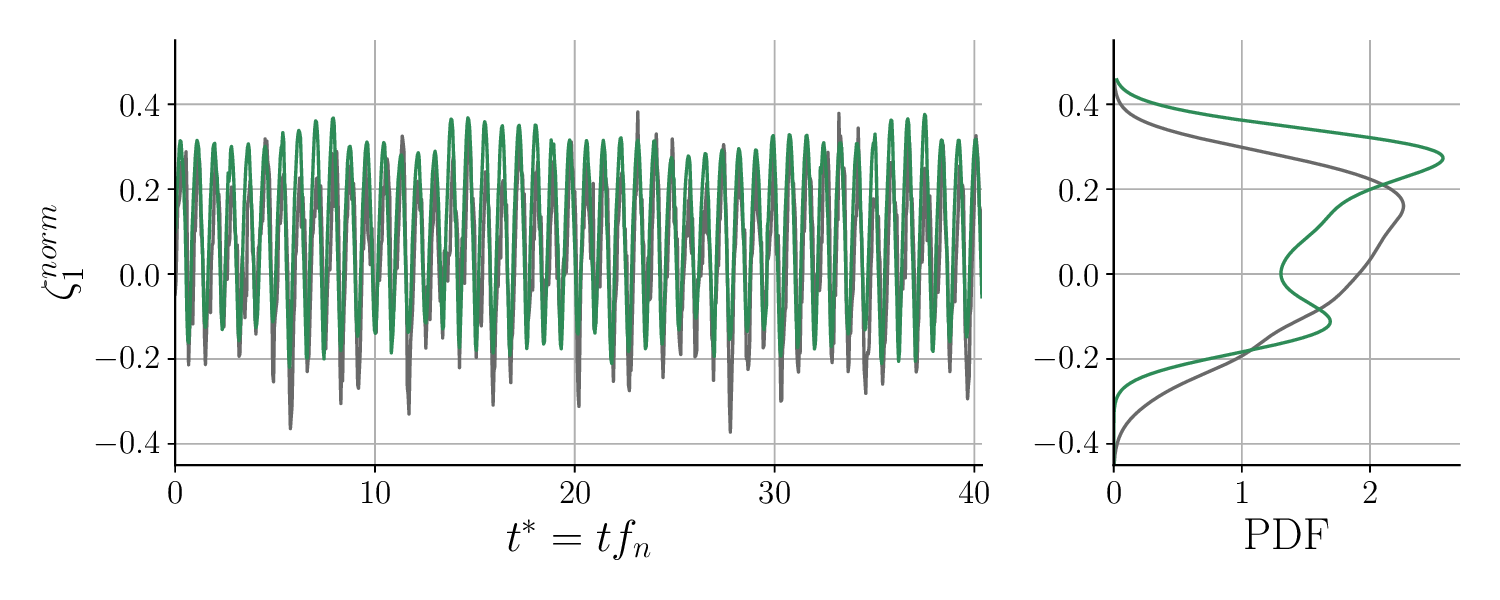}
        \includegraphics[width=\textwidth]{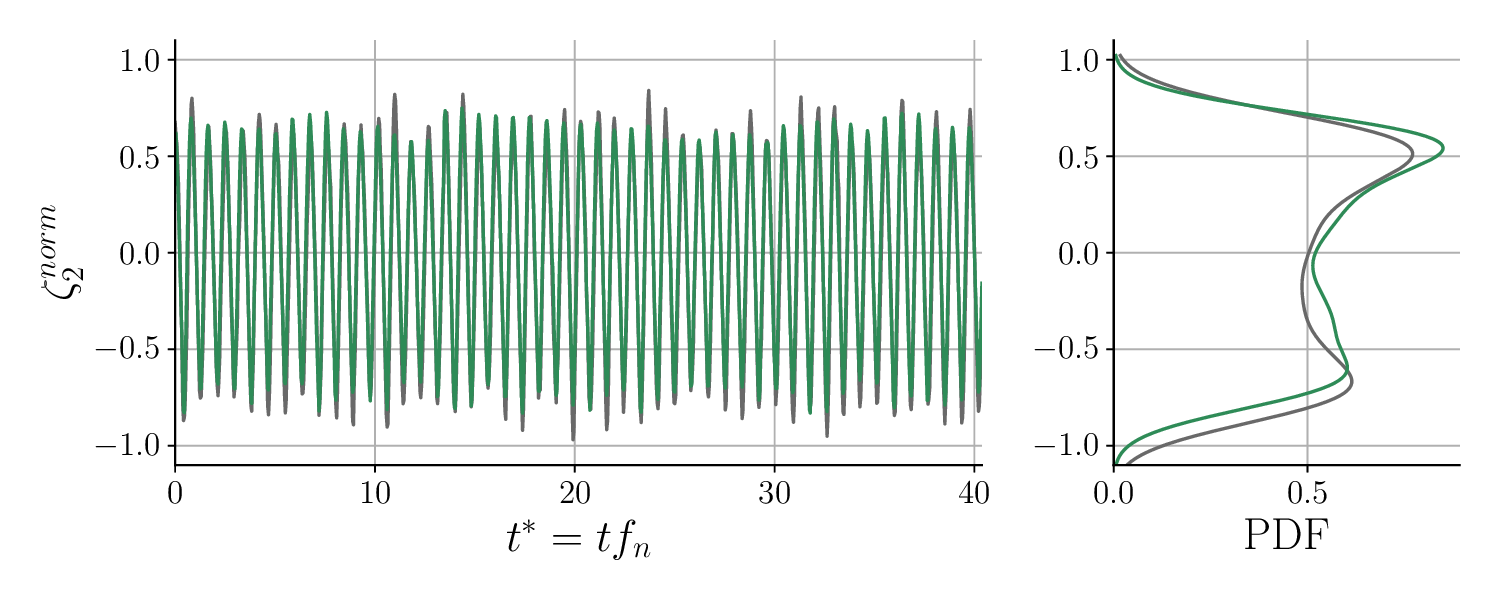}
        \includegraphics[width=\textwidth]{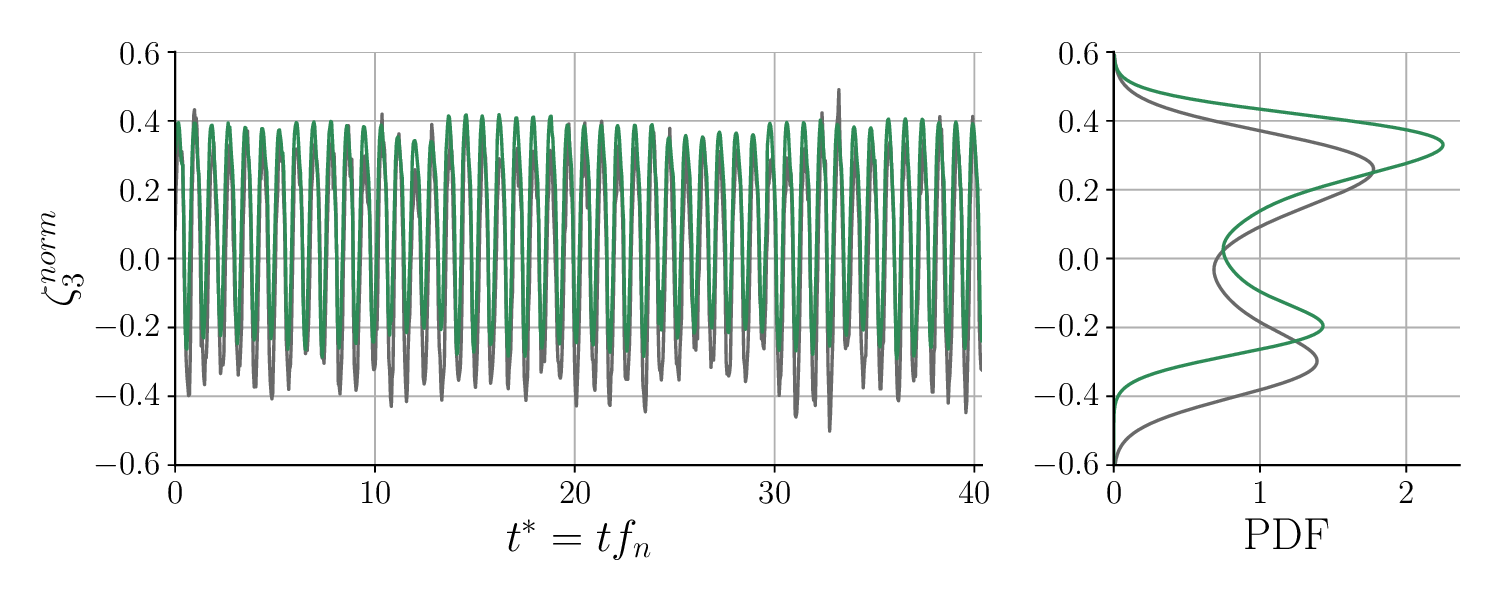}
        \subcaption{}
        \label{fig:04.2/time_series_0803}
    \end{subfigure}
    \caption{Temporal evolution comparison between target (gray) and predicted (green) latent variables for (a) upper branch ($U^*=5.56$) and (b) lower branch ($U^* = 8.03$). Time is normalised by the cylinder natural oscillation frequency ($t^* = t f_n$). Each panel presents time series data (left) and corresponding probability density functions (right).}
    \label{fig:04.2/combined_time_series}
\end{figure}

The different dynamical behaviors of the two operating conditions are well captured in the temporal and statistical characterization of the latent trajectories (Figure~\ref{fig:04.2/time_series_0556} and Figure~\ref{fig:04.2/time_series_0803}).

The temporal evolution of all three latent variables in the upper regime (Figure~\ref{fig:04.2/time_series_0556}, left panels) is characterized by pronounced non-harmonic features. This is most evident in $\zeta_1$, but also observed in the strong modulation of $\zeta_2$ and $\zeta_3$, which present more quasi-periodic behavior. 

The transformer model predictions demonstrate adequate phase synchronization across all dimensions with a mean absolute percentage error of 8.87\%. Discrepancies in amplitude reconstruction are evident, with the model generally underestimating peak magnitudes observed in the target trajectory.

The probability density function (PDF) analysis (Figure~\ref{fig:04.2/time_series_0556}, right panels) reveals distinct shapes for each latent variable. The $\zeta_2$ component manifests a bimodal distribution with approximately symmetric peaks, which is characteristic of dynamics with a dominant sinusoidal component. In contrast, $\zeta_1$ exhibits significant positive skewness (right-tail asymmetry), while $\zeta_3$ demonstrates pronounced negative skewness (left-tail asymmetry). These complementary asymmetric distributions reflect the observed phase-modulated oscillatory dynamics of this regime, consistent with the irregular vortex dynamics observed experimentally by Khalak and Williamson \cite{khalak1999motions}. 

The transformer predictions demonstrate effective shape preservation; however, they also manifest systematically narrower PDFs. This variance deficit indicates a limitation in the transformer's capacity to fully capture the variability of the test subset, particularly affecting representation of non-Gaussian statistical features as the asymmetric tails in $\zeta_1$ and $\zeta_3$.

The lower-branch latent variables evolution (Figure~\ref{fig:04.2/time_series_0803}, left panels) exhibits modifications in the flow dynamics. While the upper and lower branches are well characterised by pairs of counter-rotating vortices, the wake dynamics is different. Increased regularity is observed across all latent variables, especially $\zeta_2$ and $\zeta_3$, which present reduced amplitude modulations. This weakly modulated nature is directly related to the stable 2P shedding observed for this regime \cite{khalak1999motions}, which represents a more coherent, phase-stabilized flow organization. Despite this regime transition, the transformer maintains robust phase synchronization across all dimensions with a mean absolute percentage error of 9.92\%. Amplitude discrepancies persist and, in some cases, increase moderately compared to the upper-branch case; this is most notable for $\zeta_1$, where the attenuation of predicted amplitudes relative to the target trajectory is most pronounced.

The flow reorganization is most evident in the lower branch latent variable distributions (Figure~\ref{fig:04.2/time_series_0803}, right panels). Compared to the upper branch, $\zeta_2$ and $\zeta_3$ display statistical homogenization, both exhibiting similar bimodal PDFs. These distributions are characteristic of the two dominant oscillation modes, consistent with the phase portrait shown in the following section analysis (Figure~\ref{fig:04.2/phase_portrait_zeta23_0803}) and the high vortex periodicity observed by Khalak and Williamson \cite{khalak1999motions}. The $\zeta_1$ PDF also exhibits homogenization, with considerably reduced skewness compared to the upper branch, indicating weakened non-harmonic phenomena. 

The transformer successfully reproduces the qualitative bimodal structure of $\zeta_2$ and $\zeta_3$ PDFs. However, the predicted $\zeta_1$ distribution exhibits bimodal behavior not observed in the target latents, suggesting model limitations in capturing the statistical properties of this component. As in the upper-branch case, reduced variance is observed across all predictions.

\subsubsection{Phase Portraits and Attractor Topology}
\label{sec:04_2/phase_portrait}

\begin{figure}[h]
    \centering
    \begin{minipage}{\textwidth}
        \centering
        \begin{subfigure}{0.4\textwidth}
            \includegraphics[width=\textwidth]{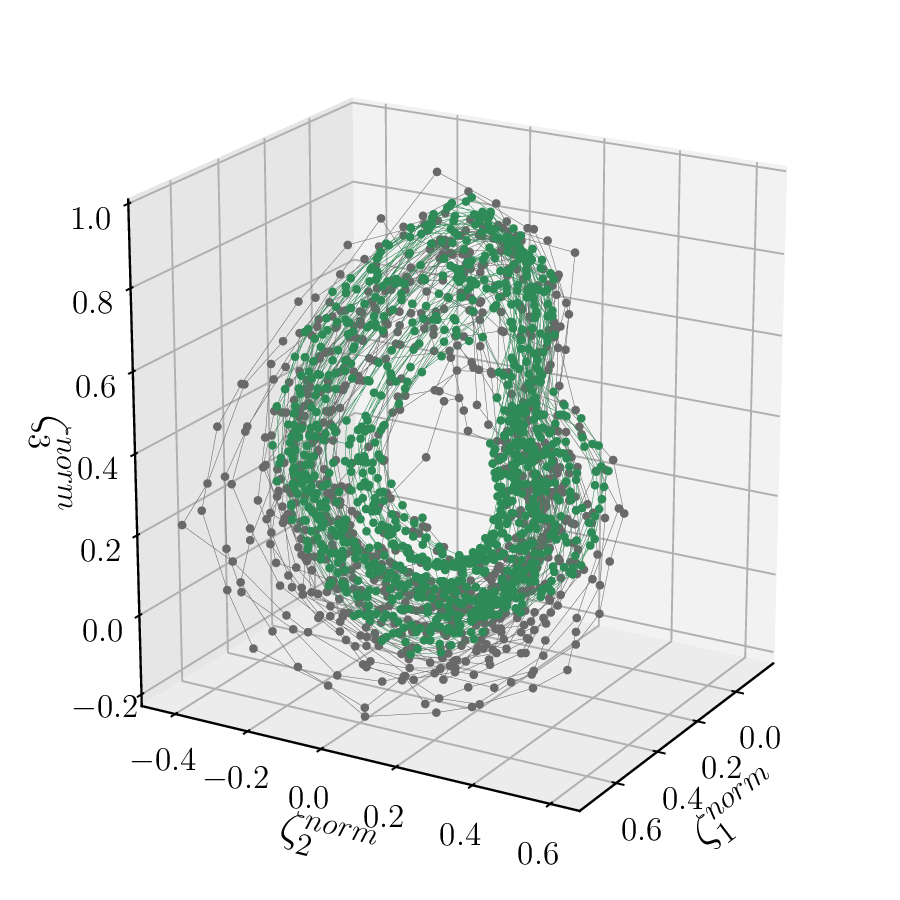}
            \caption{}
            \label{fig:04.2/phase_portrait_3d_0556}
        \end{subfigure}
    \end{minipage}
    
    \vspace{0.5cm} % Add some vertical space between the rows
    
    \begin{minipage}{\textwidth}
        \centering
        \begin{subfigure}{0.33\textwidth}
            \includegraphics[width=\textwidth]{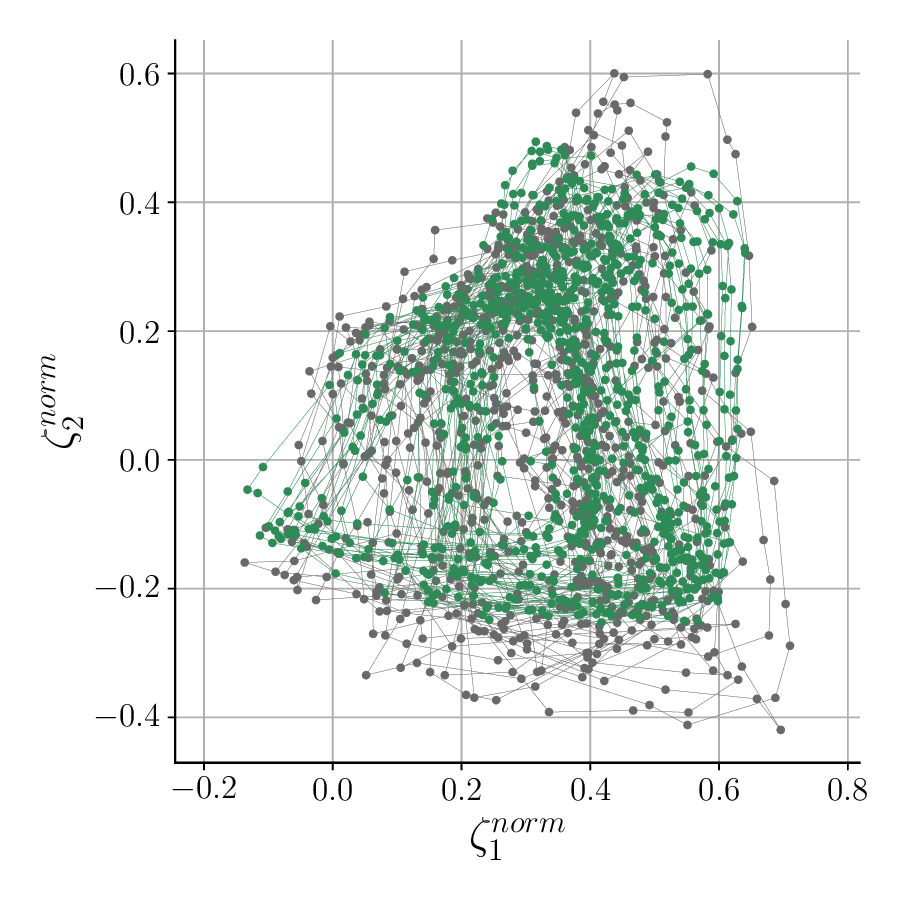}
            \caption{}
            \label{fig:04.2/phase_portrait_zeta12_0556}
        \end{subfigure}
        \hfill
        \begin{subfigure}{0.33\textwidth}
            \includegraphics[width=\textwidth]{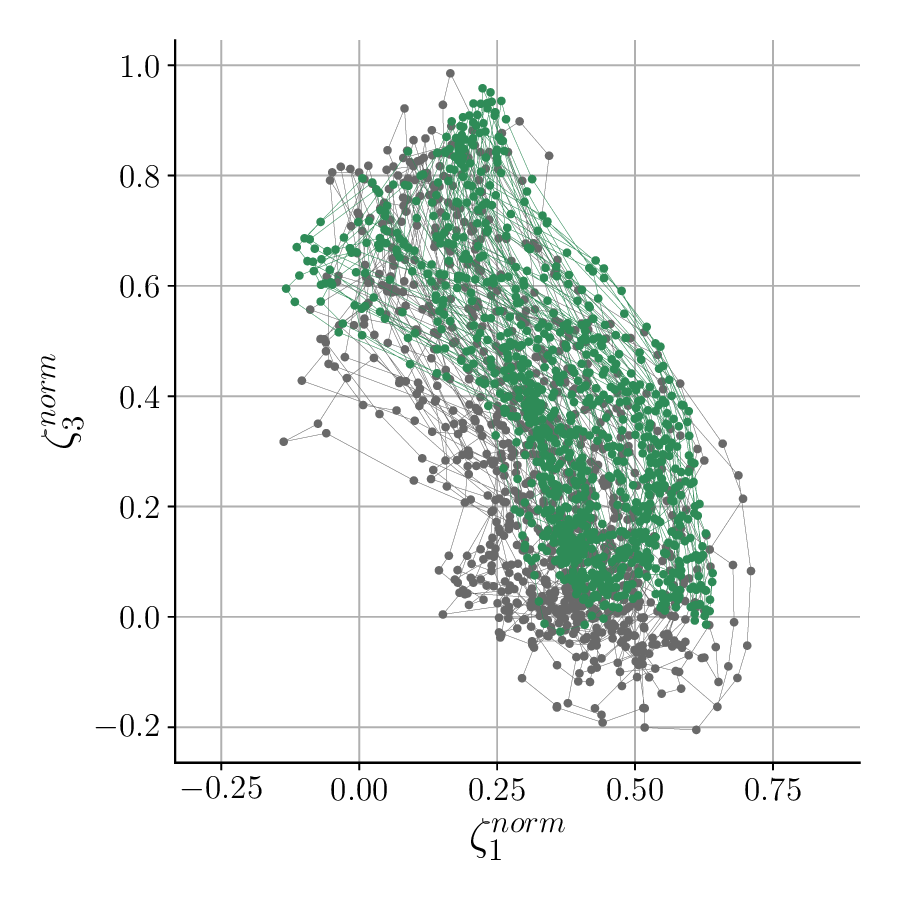}
            \caption{}
            \label{fig:04.2/phase_portrait_zeta13_0556}
        \end{subfigure}
        \hfill
        \begin{subfigure}{0.33\textwidth}
            \includegraphics[width=\textwidth]{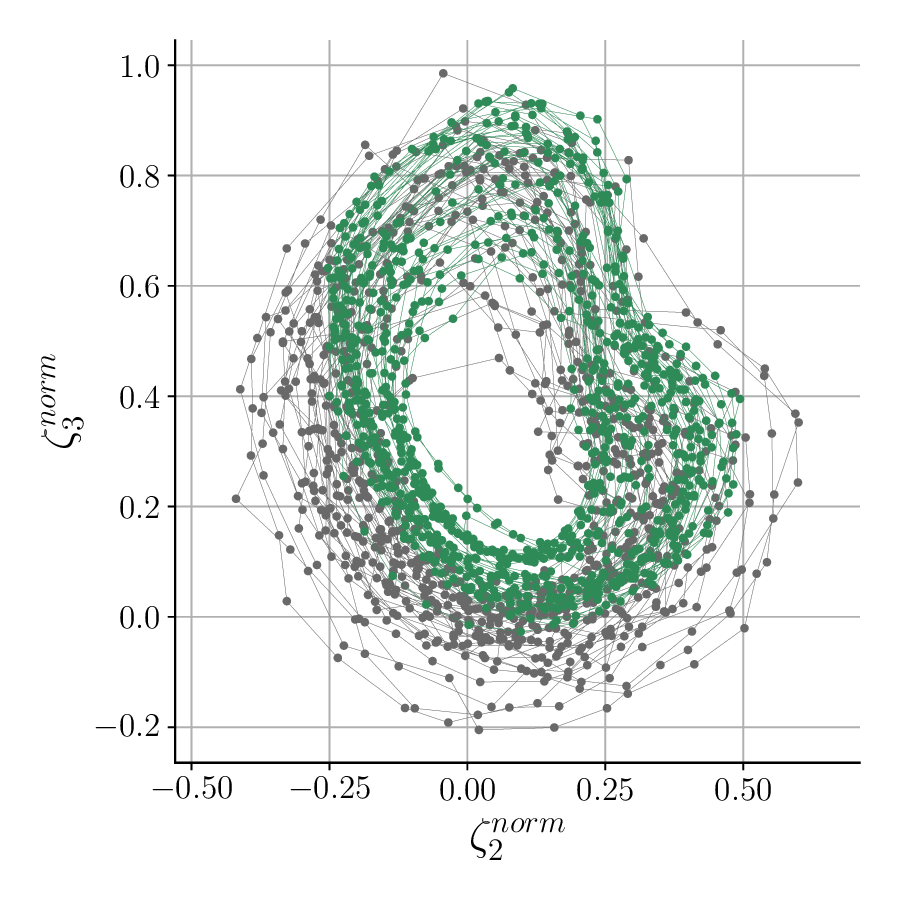}
            \caption{}
            \label{fig:04.2/phase_portrait_zeta23_0556}
        \end{subfigure}
    \end{minipage}

    \vspace{0.5cm} % Add some vertical space between the rows
    \caption{Phase portraits comparing target (gray) and predicted (green) latent-space trajectories of the upper branch regime ($U^* = 5.56$). (a) Three-dimensional representation. (b) $\zeta_1$-$\zeta_2$ projection. (c) $\zeta_1$-$\zeta_3$ projection. (d) $\zeta_2$-$\zeta_3$ projection.}
    \label{fig:04.2/phase_portraits_0556}
\end{figure}

\begin{figure}[h!]
    \centering
    \begin{minipage}{\textwidth}
        \centering
        \begin{subfigure}{0.4\textwidth}
            \includegraphics[width=\textwidth, trim=0cm 1.5cm 0.8cm 3.5cm, clip]{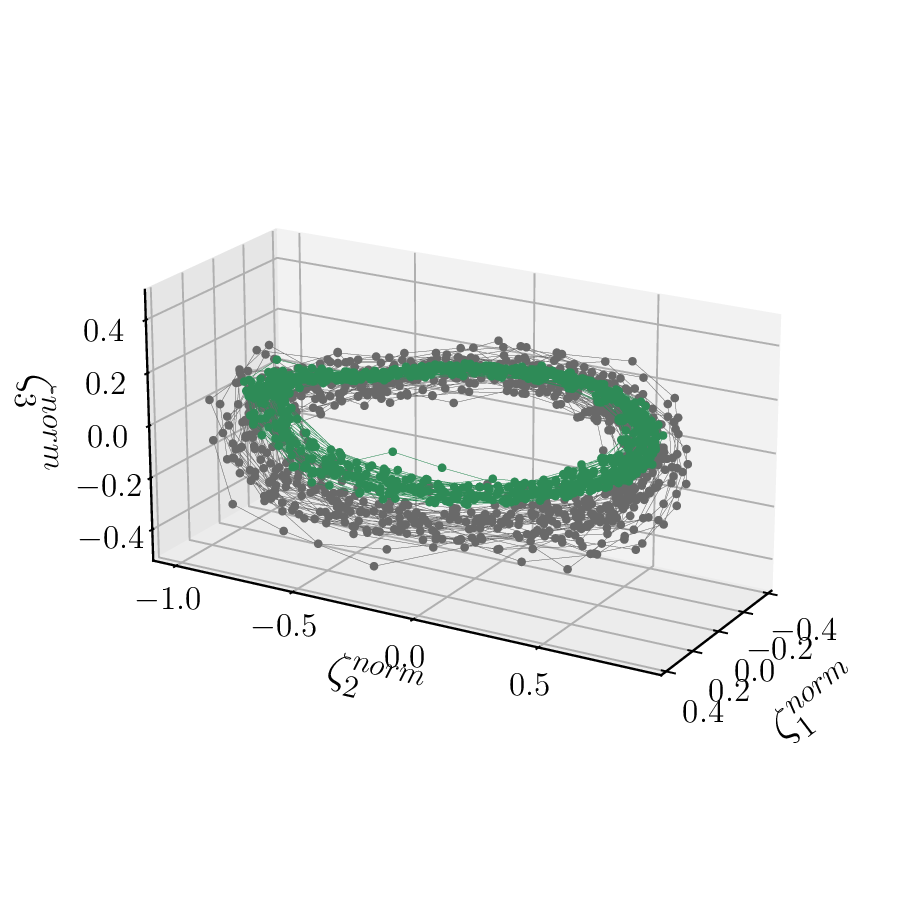}
        \caption{}
        \label{fig:04.2/phase_portrait_3d_0803}
        \end{subfigure}
    \end{minipage}
    
    \vspace{0.5cm} % Add some vertical space between the rows
    
    \begin{minipage}{\textwidth}
        \centering
        \begin{subfigure}{0.33\textwidth}
            \includegraphics[width=\textwidth]{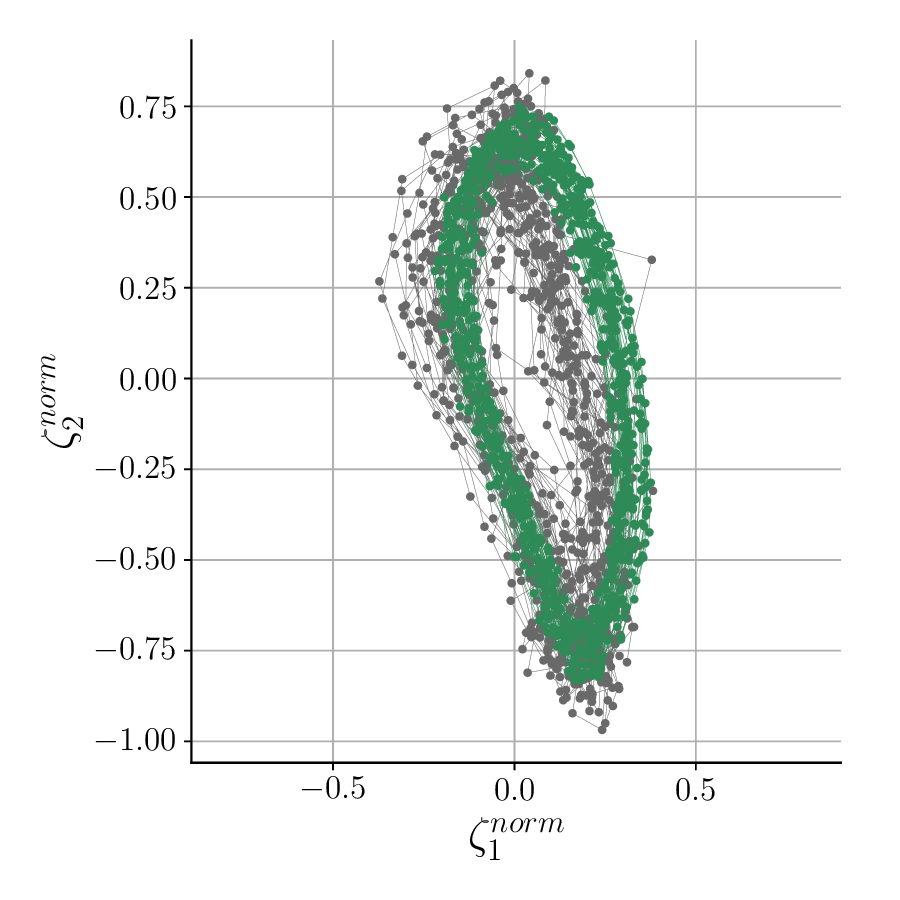}
            \caption{}
            \label{fig:04.2/phase_portrait_zeta12_0803}
        \end{subfigure}
        \hfill
        \begin{subfigure}{0.33\textwidth}
            \includegraphics[width=\textwidth]{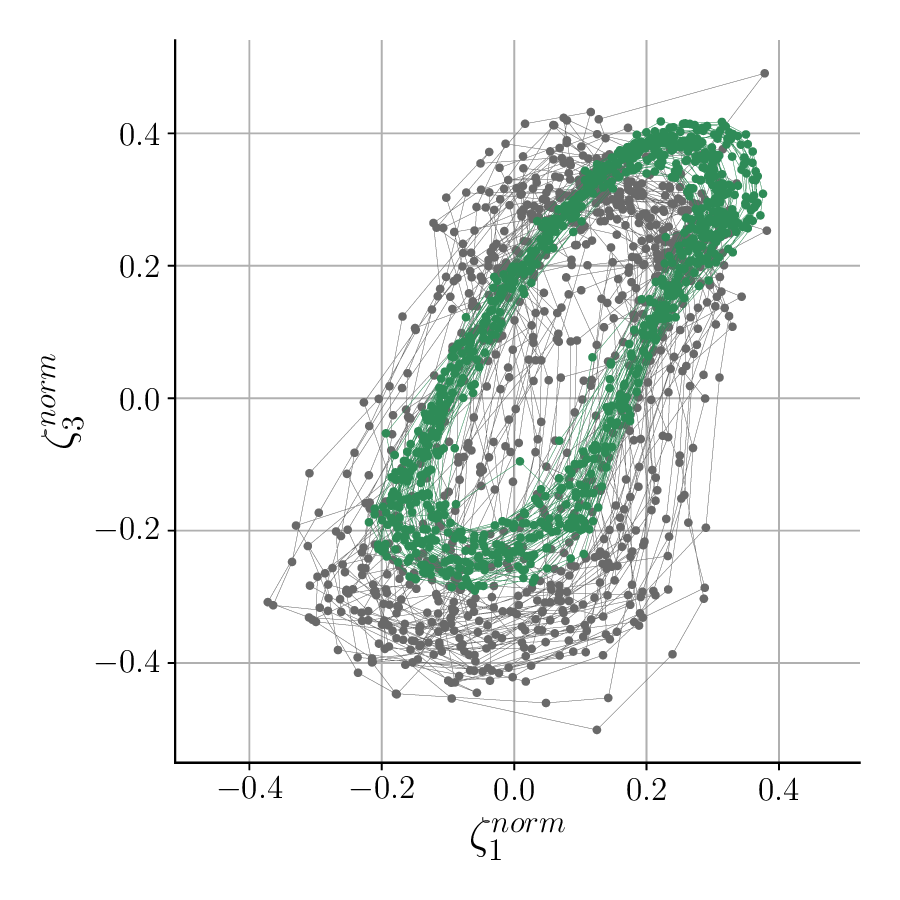}
            \caption{}
            \label{fig:04.2/phase_portrait_zeta13_0803}
        \end{subfigure}
        \hfill
        \begin{subfigure}{0.33\textwidth}
            \includegraphics[width=\textwidth]{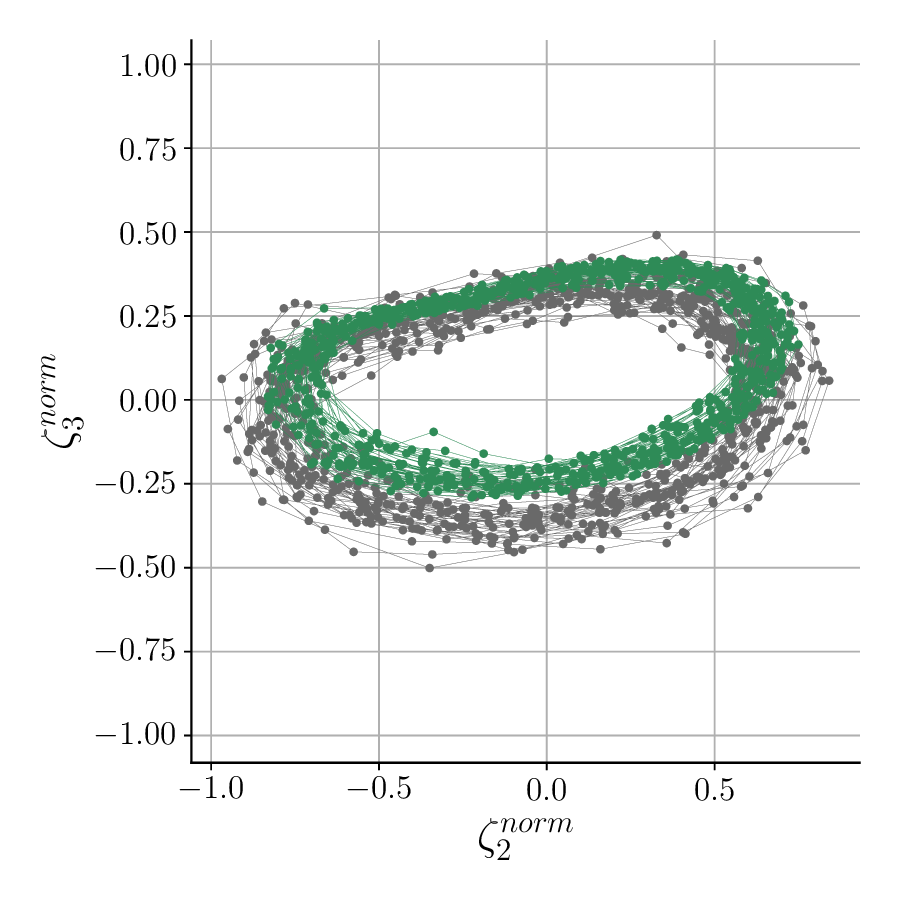}
            \caption{}
            \label{fig:04.2/phase_portrait_zeta23_0803}
        \end{subfigure}
    \end{minipage}

    \vspace{0.5cm} % Add some vertical space between the rows
    \caption{Phase portraits comparing target (gray) and predicted (green) latent-space trajectories of the lower branch regime ($U^* = 8.03$). (a) Three-dimensional representation. (b) $\zeta_1$-$\zeta_2$ projection. (c) $\zeta_1$-$\zeta_3$ projection. (d) $\zeta_2$-$\zeta_3$ projection.}
    \label{fig:04.2/phase_portraits_0803}
\end{figure}

Phase-space portraits are used to assess further the transformer's predictive performance in characterizing the invariant manifolds that govern system behavior. Figures~\ref{fig:04.2/phase_portraits_0556}-\ref{fig:04.2/phase_portraits_0803} present comprehensive visualizations of the latent-space phase portraits for both flow regimes. The  topological  correspondence  between  predicted and target attractors is quantitatively assessed through a convex hull analysis. The convex hull $\mathcal{H}(P)$ of phase-portrait points $P = \{p_1, p_2, ..., p_n\}$ is defined as the minimal convex enclosure containing all points in $P$:

\begin{equation}
	\mathcal{H}(P) = \left\{ \sum_{i=1}^{n} \gamma_i p_i \mid \gamma_i \geq 0, \sum_{i=1}^{n} \gamma_i = 1 \right\}.
\end{equation}

\noindent The computations use Delaunay triangulation to decompose $\mathcal{H}(P)$ into simplices, with point containment determined via barycentric coordinate analysis. Volume ratios $V_{\mathrm{pred}}/V_{\mathrm{gt}}$ and containment analysis then quantify state space volume and exploration, where containment analysis measures the fraction of predicted trajectory points that lie within the target attractor's convex hull and the fraction of target points within the predicted convex hull.

The upper branch regime (Figure~\ref{fig:04.2/phase_portraits_0556}) presents a roughly annular structure with points broadly spread in the radial and axial direction. 

This ring-like structure is most evident in the $\zeta_2$-$\zeta_3$ projection, where quasi-periodic trajectories are observed. This projection shape is comparable to the one observed in the two dominant POD modes of the same branch found in literature, for example, in Janocha \textit{et al.} \cite{janocha2022large} and Riches \textit{et al}  \cite{riches2018proper}. This topological feature also aligns with the bimodal PDF of $\zeta_2$ observed in the temporal evolution analysis (Figure~\ref{fig:04.2/time_series_0556}). These observations suggest that these two latent dimensions collectively encode the principal oscillatory mechanisms governing vortex formation and shedding. The $\zeta_1$-$\zeta_2$ projection, instead, exhibits a diffuse, irregular distribution with concentrated density in the central region (Figure~\ref{fig:04.2/phase_portrait_zeta12_0556}). The pattern reflects the non-harmonic features observed in the temporal evolution of these variables, and it is comparable to the third POD mode pair found in the analysis of Riches \textit{et al} \cite{riches2018proper}, where the authors associate it with an intermittent behavior of the wake. The $\zeta_1$-$\zeta_3$ projection displays a linear relation with negative slope (Figure~\ref{fig:04.2/phase_portrait_zeta13_0556}) representing a phase opposition relationship between these two modes, in accordance with the strong anti-correlation $\rho(\zeta_1,\zeta_3) \approx -0.66$ quantified in the correlation analysis (Section \ref{sec:04_2/correlation_analysis}). This pattern is consistent with the oppositely directed skewness observed in their respective probability density functions (Figure~\ref{fig:04.2/time_series_0556}) and is not evident in linear POD analyses. The observed anti-correlation between $\zeta_1$ and $\zeta_3$ may relate with evidence of competing flow modes in the upper branch regime. Zhao et al. \cite{zhao2014chaotic} reported chaos arising from competition between distinct vortex shedding modes, while Morse and Williamson \cite{morse2009prediction} discovered overlapping vortex formation patterns in this regime. Specifically, Morse and Williamson identified intermittent switching between the standard 2P mode (typical of lower branch) and a newly discovered 2Po mode. This 2Po mode occurs at peak resonant response, the condition corresponding to the upper branch test case examined in this analysis. The strong anti-correlation, asymmetric PDF distributions, and linear phase relationship collectively suggest that $\zeta_1$ and $\zeta_3$ may capture these competing dynamical states. The intermittent behavior previously associated with $\zeta_1$-$\zeta_2$ projection further supports this interpretation, potentially reflecting the mode-switching dynamics observed by Morse and Williamson. Therefore, this may represents a form of nonlinear mode interaction that the ML approach captures while linear decomposition methods cannot detect. However, further investigation would be needed to establish the definitive physical significance of this relationship.

The transformer predictions align with the target attractor shape but exhibit volumetric contraction. Volume ratio analysis indicates the predicted attractor occupies 55\% of the target manifold volume ($V_{\mathrm{pred}}/V_{\mathrm{gt}} = 0.55$), consistent with the peak underestimation observed in temporal evolution analysis. Containment analysis reveals that 98.4\% of predicted attractor points lie within the target attractor hull, while only 51.4\% of target points are contained within the predicted hull. This asymmetric containment pattern indicates the model successfully captures core attractor structure but fails to reproduce the full extent of system variability, as also observed in the PDF analysis.

The lower branch manifold (Figure~\ref{fig:04.2/phase_portraits_0803}) presents well defined annular structures. The points spread in the axial and radial directions is more constrained with respect to the upper branch regime. 

This ring structure is consistent with the highly periodic vortex dynamics observed by Khalak and Williamson \cite{khalak1999motions} and with the increased regularity observed in the trajectories temporal evolution (Figure~\ref{fig:04.2/time_series_0803}, left). As in the upper branch, this annular shape is most evident in the $\zeta_2$-$\zeta_3$ projection, consistent with bimodal PDFs observed in the statistical analysis (Figure~\ref{fig:04.2/time_series_0803}, right). The projection shape is comparable to the one of the dominant POD mode pair found in Janocha \textit{et al.} \cite{janocha2022large}, who also report a reduced radial point spread of the modes compared to the upper branch, as observed in this work. As for the upper branch, these modes are linked to the principal oscillatory mechanisms. Instead, $\zeta_1$-$\zeta_2$ and $\zeta_1$-$\zeta_3$ projections appear as elongated elliptical structures oriented in opposed diagonal directions. The orientation of these structures encode specific latent variables relationship, that will be further studied in Section \ref{sec:04_2/correlation_analysis}. These elliptical patterns are not evident in any of the linear decomposition study found in the literature and is therefore believed to be linked to nonlinear mode interactions captured by the ML approach, as also suggested by the irregular evolution of $\zeta_1$ observed in the time analysis. 

The transformer predictions align with the target attractor shape but exhibit greater volumetric contraction than the upper branch regime. The predicted attractor occupies only 23\% of the target manifold volume ($V_{\mathrm{pred}}/V_{\mathrm{gt}} = 0.23$). As for the upper branch, containment analysis shows that almost all predicted points (97.5\%) lie within the target space region. However, only 28.7\% of target points lie within the predicted region. Therefore, consistent with the reduced variability observed in PDF analysis, the transformer predictions fail to represent the full system extent but successfully capture the core attractor.

\subsubsection{Correlation Structure}
\label{sec:04_2/correlation_analysis}

\begin{figure}[h!t]
	\centering
	\includegraphics[width=0.45\textwidth]{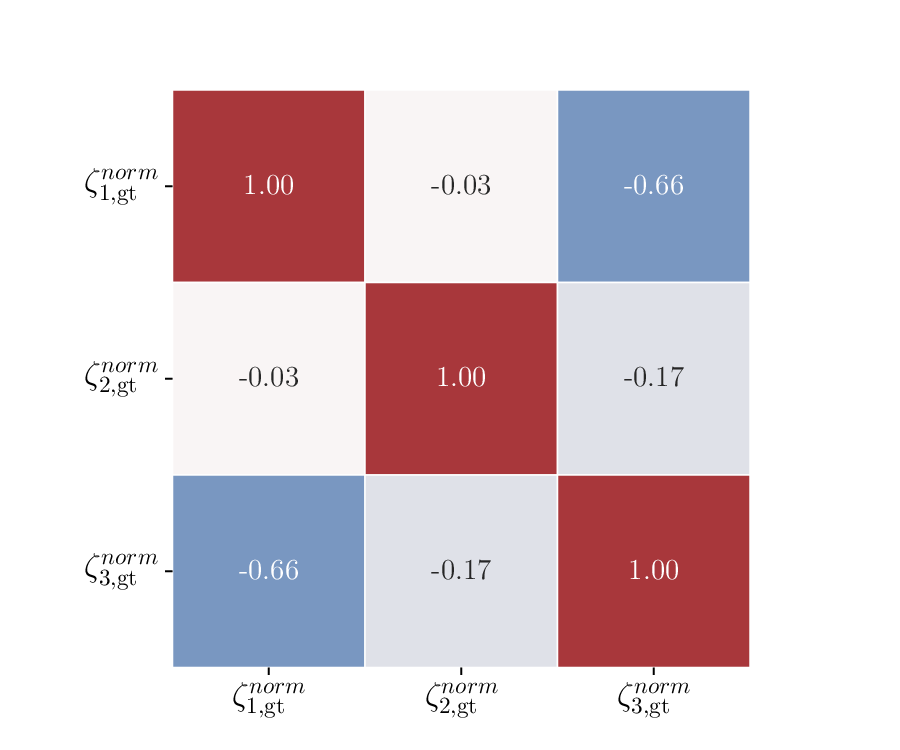}
	\includegraphics[width=0.45\textwidth]{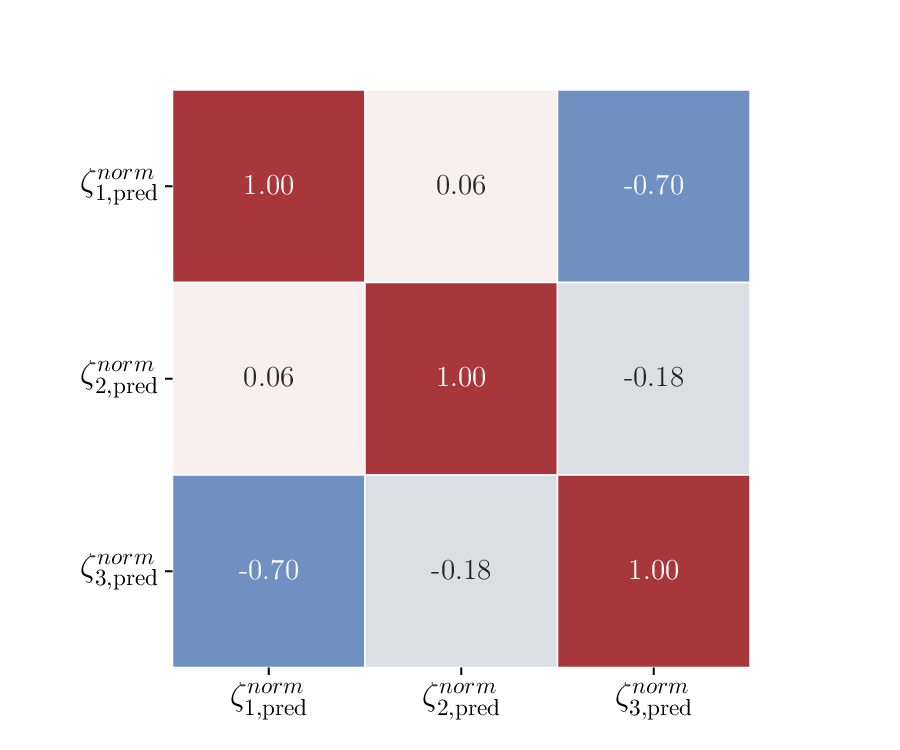}\\
	\vspace{0.3cm}
	\includegraphics[width=0.45\textwidth]{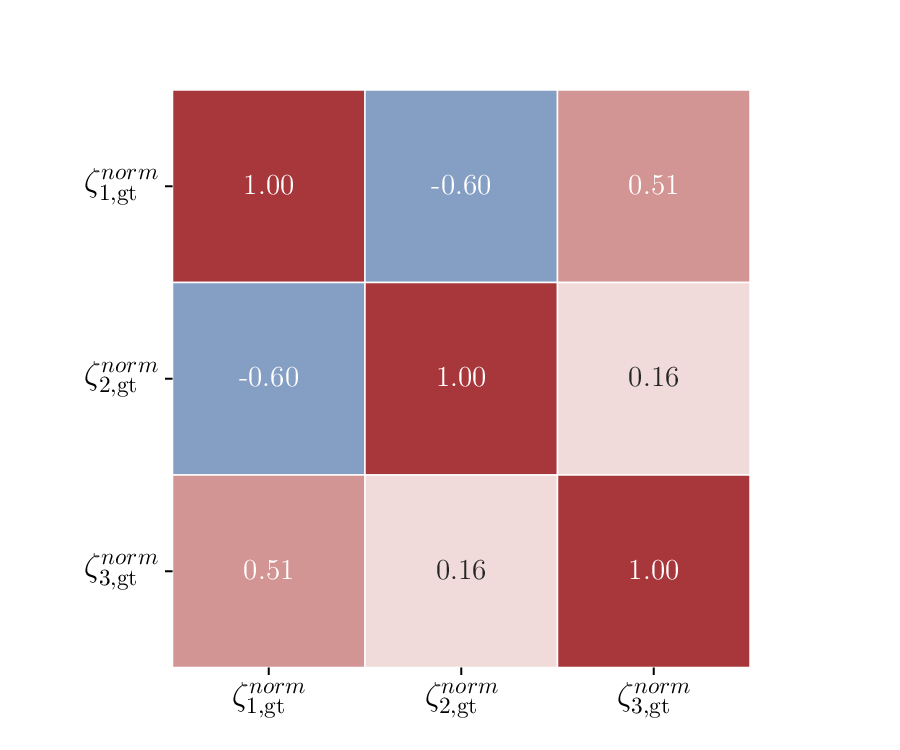}
	\includegraphics[width=0.45\textwidth]{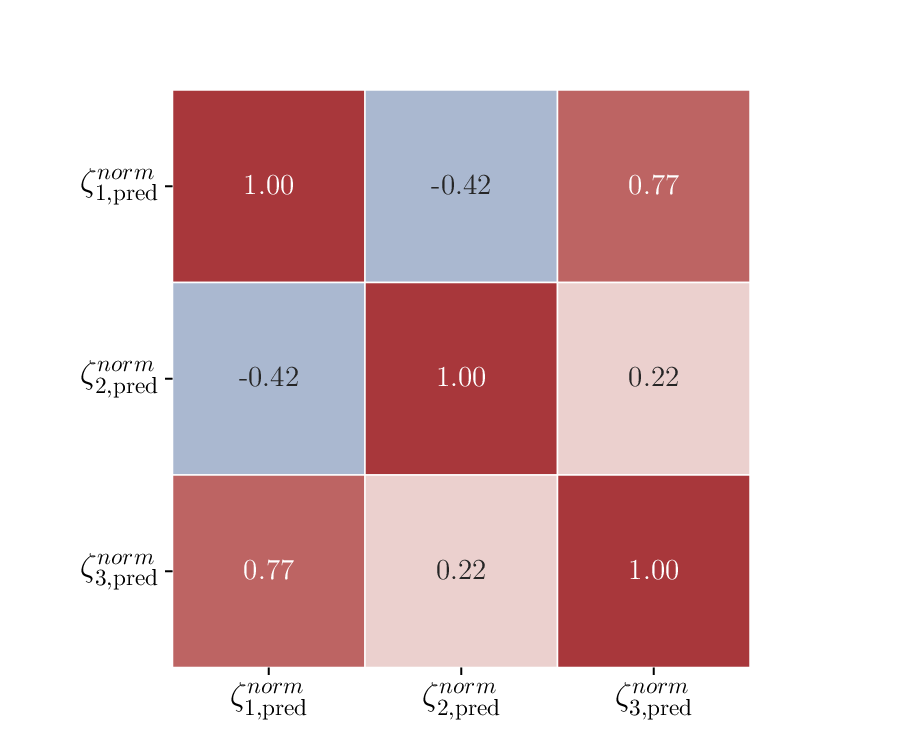}
	\caption{Left: Target  correlation matrices for the latent variables. Right: Prediction correlation matrices. Top row: upper branch regime ($U^* = 5.56$). Bottom row: lower branch regime ($U^* = 8.03$).}
	\label{fig:04_2/correlation_analysis}
\end{figure}

Correlation analysis is performed to characterize the relationships between latent variables and assesses the transformer's capacity to reproduce these dependencies. Figure~\ref{fig:04_2/correlation_analysis} presents correlation matrices for both flow regimes. Each matrix element is defined by the Pearson correlation coefficient:

\begin{equation}
 	\rho(\zeta_i, \zeta_j) = \frac{\mathbb{E}[(\zeta_i - \mu_{\zeta_i})(\zeta_j - \mu_{\zeta_j})]}{\sigma_{\zeta_i} \sigma_{\zeta_j}},
\end{equation}

\noindent where $\mu_{\zeta_i}$ is the ensemble mean of the $i$-th latent variable and $\sigma_{\zeta_i}$ is its standard deviation.

In the upper-branch regime, the target correlation matrix (Figure~\ref{fig:04_2/correlation_analysis}, top left) reveals a near-diagonal structure with predominantly uncorrelated latent variables, except for a single dominant anti-correlation $\rho(\zeta_1, \zeta_3) \approx -0.66$. This structure indicates that the $\beta$-VAE-GAN encoder has disentangled the flow physics into largely independent latent features. The pronounced anti-correlation corresponds directly to the linear relationship observed in the $\zeta_1$-$\zeta_3$ phase projection (Figure~\ref{fig:04.2/phase_portrait_zeta13_0556}), signifying phase opposition between these complementary modes.

The transformer prediction correlation matrix closely reproduces the target structure, with most elements showing only minor quantitative departures from target values. For the $\zeta_1$-$\zeta_2$ pair, the correlation sign changes from negative $\rho(\zeta_1, \zeta_2) = -0.03$ to positive $\rho(\zeta_1, \zeta_2) = 0.06$. However, both correlations are negligible, confirming that these variables remain essentially uncorrelated in both cases. 

The lower branch regime reveals different mode relationships. While the two dominant oscillation modes $\zeta_2-\zeta_3$ presents correlation values $\rho(\zeta_2, \zeta_3) \approx 0.16$ comparable to the upper branch, two additional correlated pairs emerge: $\zeta_1$-$\zeta_2$ and $\zeta_1$-$\zeta_3$. These pairs exhibit opposite signed correlations, consistent with the opposing orientations of their elliptical phase portrait structures observed in Figure~\ref{fig:04.2/phase_portraits_0803}. The elliptical orbits also suggest oscillatory behavior. Under the simplified assumption of equal-amplitude sinusoidal oscillators, correlations approximately relate to phase differences as:

\begin{equation}
	\rho(\zeta_i, \zeta_j) \approx \cos(\Delta\phi_{ij}).
\end{equation}

\noindent Following this analogy, $\zeta_1$-$\zeta_2$ exhibits a phase offset of approximately $\Delta\phi_{12} \approx -53^{\circ}$, while $\zeta_1$-$\zeta_3$ shows $\Delta\phi_{13} \approx 60^{\circ}$. These nearly complementary phase relationships suggest $\zeta_1$ may mediate coupling between the two dominant oscillatory modes. However, the assumption made may be an over-simplification of the more rich interaction between these modes and further investigation is required to understand the physical mechanism encoded by this variable.

The transformer prediction correlation matrices show larger departures from the target values than what observed in the upper branch regime. However, correlation signs remain unchanged, indicating preservation of the fundamental variables relationships. These departures are consistent with the higher prediction errors and the $\zeta_1$ PDF mismatch discussed in the temporal and statistical analysis. The larger deviations suggest the transformer encounters greater difficulty accurately reproducing this flow regime, though it successfully captures the underlying mode coupling structure. This increased prediction challenge for the lower branch is also consistent with the greater volumetric contraction observed in phase space analysis.

\subsection{Combining the transformer and the $\beta$-VAE-GAN decoder: VIVALDy inference}
\label{sec:04.3}

Having established the individual performances of the $\beta$-VAE-GAN decoder in flow field reconstruction (Section \ref{sec:04_1}) and the bidirectional transformer in latent-space trajectory prediction (Section \ref{sec:04_2}), this section now integrates these two models to form the complete VIVALDy inference framework and evaluates its end-to-end performance in predicting velocity fields of the test set  $\mathcal{D}^{\mathrm{Test}}$ from cylinder displacement $y_{\mathrm{cyl}}$.

\subsubsection{Reconstruction Accuracy and Distribution Alignment}

Table \ref{tab:04_3/vivaldy_metrics_transposed} quantifies VIVALDy reconstruction accuracy and distributional alignment performance using the metrics introduced in Section \ref{sec:04_1/evaluation_metrics}. Performance is evaluated relative to a reference case using latent variables encoded from the ground truth PIV snapshots, thereby isolating the transformer's contribution to overall system performance. For brevity, $\text{NRMSE}_u$ and $W_u$ denote the respective metrics for the $u$-component, with analogous notation for the $v$-component.

\begin{table}[h!]
    \centering
    \begin{tabular}{lcccc}
    \toprule\toprule
        Flow Regime & $\text{NRMSE}_{u}$ & $\text{NRMSE}_{v}$ & $W_u$ & $W_v$ \\
        \midrule
        Initial     & \textcolor{red}{0.0960 (+55.33\%)} & \textcolor{red}{0.1912 (+106.0\%)}    & 0.0033 (+13.72\%)    & 0.0026 (+18.18\%) \\
        Upper       & 0.1023 (+10.47\%)                  & 0.1211 (+16.00\%)                     & 0.0052 (+18.18\%)    & 0.0039 (-18.75\%) \\
        Transition  & 0.1180 (+18.83\%)                  & \textcolor{red}{0.1483 (+53.20\%)}    & 0.0050 (-16.66\%)    & \textcolor{red}{0.0075 (+53.06\%)} \\
        Lower       & 0.1042 (+24.49\%)                  & 0.1185 (+34.66\%)                     & 0.0045 (-4.255\%)    & \textcolor{red}{0.0079 (+64.58\%)} \\
        Asynchrony  & 0.0811 (+7.846\%)                  & 0.1204 (+22.11\%)                     & 0.0086 (-1.149\%)    & 0.0083 (+6.410\%) \\
    \bottomrule\bottomrule
    \end{tabular}
    \vspace{0.5cm}
    \caption{Quantitative metrics for VIVALDy inference performance across different flow regimes. Percentages in parentheses indicate changes relative to the reference case where ground truth encoded latent variables are fed into the same decoder compared to using transformer-predicted latents. Red values highlight a significant worsening in performance.}
    \label{tab:04_3/vivaldy_metrics_transposed}
\end{table}

NRMSE values are mostly higher than the reference case, however their magnitudes remain low across all flow regimes. The transformer-predicted latent variables introduce accuracy losses in specific regimes. The initial regime shows increases of 55.33\% for $\text{NRMSE}_{u}$ and 106\% for $\text{NRMSE}_{v}$, likely due to the nearly static cylinder motion in this operating condition providing a low signal-to-noise ratio for the transformer input. The transition regime exhibits increases of 18.83\% for $\text{NRMSE}_{u}$ and 53.20\% for $\text{NRMSE}_{v}$, which can be linked to this regime's absence from the training set, requiring greater model generalizability. Instead, the lower branch regime show increases of 24.49\% for $\text{NRMSE}_{u}$ and 34.66\% for $\text{NRMSE}_{v}$, which are consistent with the latent space analysis findings, where a mismatch for the $\zeta_1$ PDF prediction and a phase space volumetric contractions were observed. Despite these increases, the model demonstrates robustness with $\text{NRMSE}_{u}$ values remaining below 0.12 and $\text{NRMSE}_{v}$ values below 0.20 across all regimes.

It is observed that values of $\text{NRMSE}_{u}$ are generally lower than $\text{NRMSE}_{v}$. This trend is attributed to the statistical properties of the flow: the $u$-component is dominated by a stationary mean profile (the wake deficit), whereas the v-component is characterized by zero-mean fluctuations. Since the stationary mean profile represents a stable, large-scale feature, it is easier for the model to learn it, resulting in higher reconstruction fidelity for $u$.

For Wasserstein distances, the results reveal asymmetric performance between velocity components. The $u$-component distributions generally show modest improvements or degradations. In contrast, the $v$-component exhibits distributional misalignments, particularly in the transition (+53.06\%) and lower (+64.58\%) regimes. This asymmetry suggests the transformer more effectively captures streamwise flow statistics, possibly due to stronger coupling between cylinder displacement and streamwise velocity fluctuations.

\subsubsection{Phase Averaged Flow Visualization}

Phase-averaged flow visualizations complement the quantitative NRMSE results. Various signals are used in the literature to define reference phases, such as pressure signals \cite{perrin2007obtaining} or transverse velocity at a point in the near wake \cite{cagney2013wake}. Here, phase averaging is based on the cylinder position, following O'Neill \textit{et al.} \cite{o2021wake}. This approach requires computing the Hilbert transform of the cylinder displacement signal to define the instantaneous phase angle. Each snapshot is then assigned to a phase bin based on its corresponding time instant, and averaged fields are computed for each bin. Following O'Neill \textit{et al.} \cite{o2021wake}, 36 bins are used.  The obtained phase-averaged fields represent the coherent velocity structures that define the dominant wake topology.

\begin{figure}[h!]
    \centering
    \includegraphics[width=0.45\textwidth, trim=0cm 3cm 0cm 3.5cm, clip]{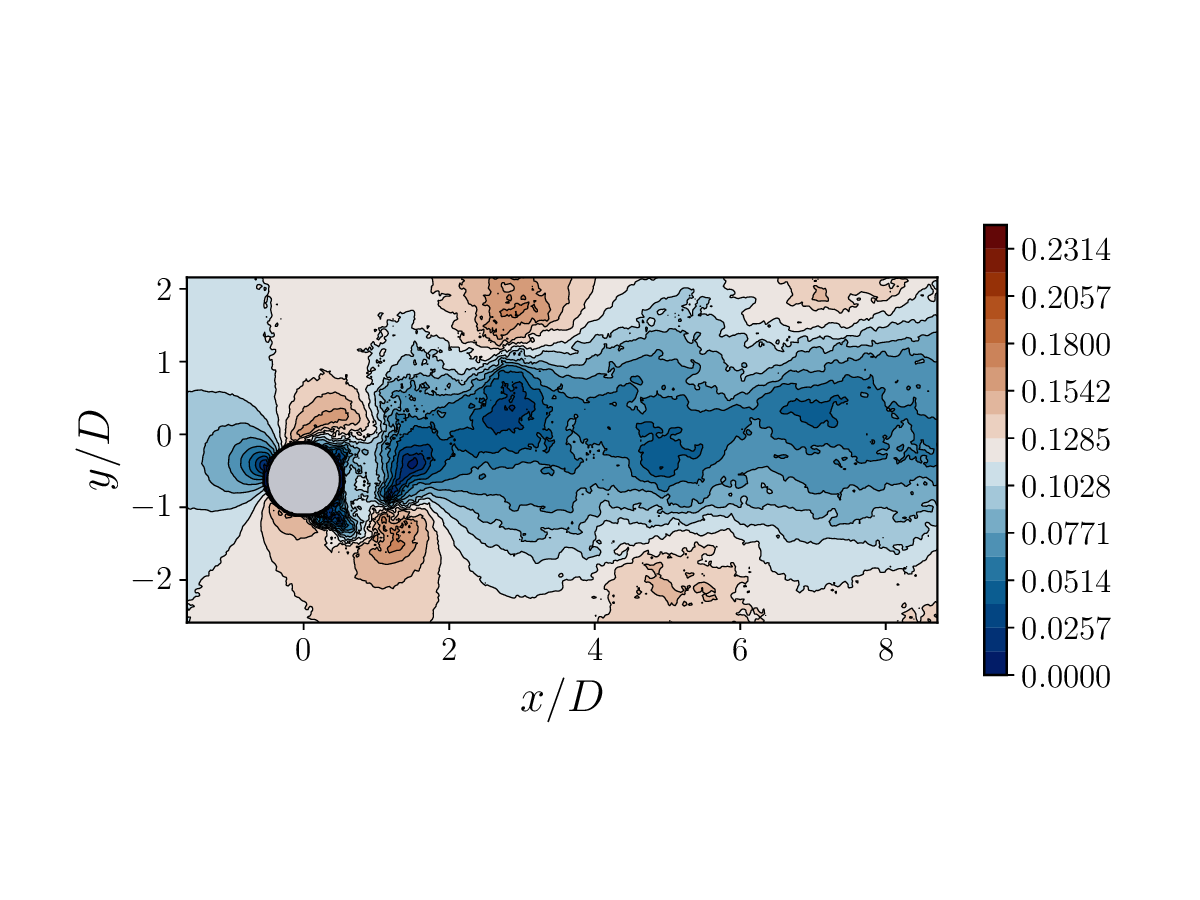}
    \includegraphics[width=0.45\textwidth, trim=0cm 3cm 0cm 3.5cm, clip]{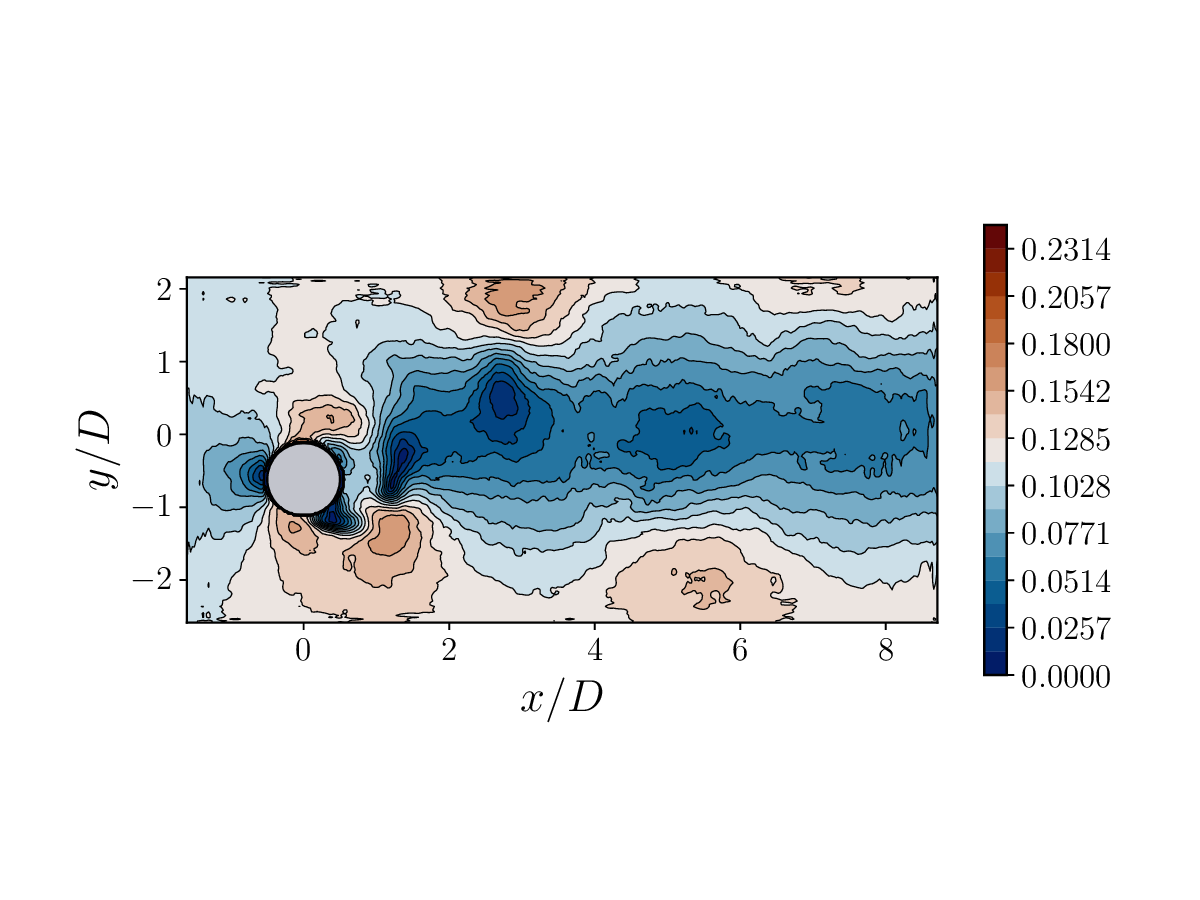}\\
    \vspace{3mm}
    \includegraphics[width=0.45\textwidth, trim=0cm 3cm 0cm 3.5cm, clip]{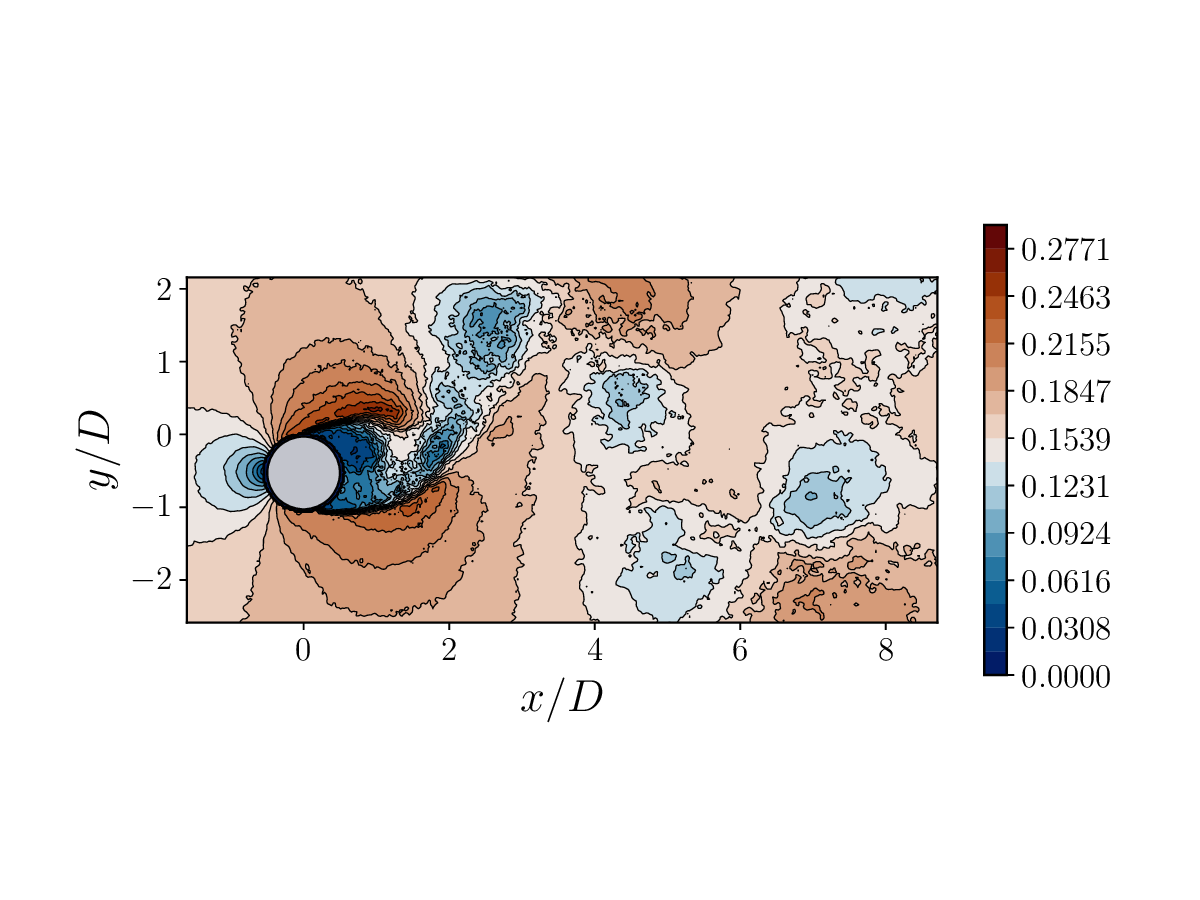}
    \includegraphics[width=0.45\textwidth, trim=0cm 3cm 0cm 3.5cm, clip]{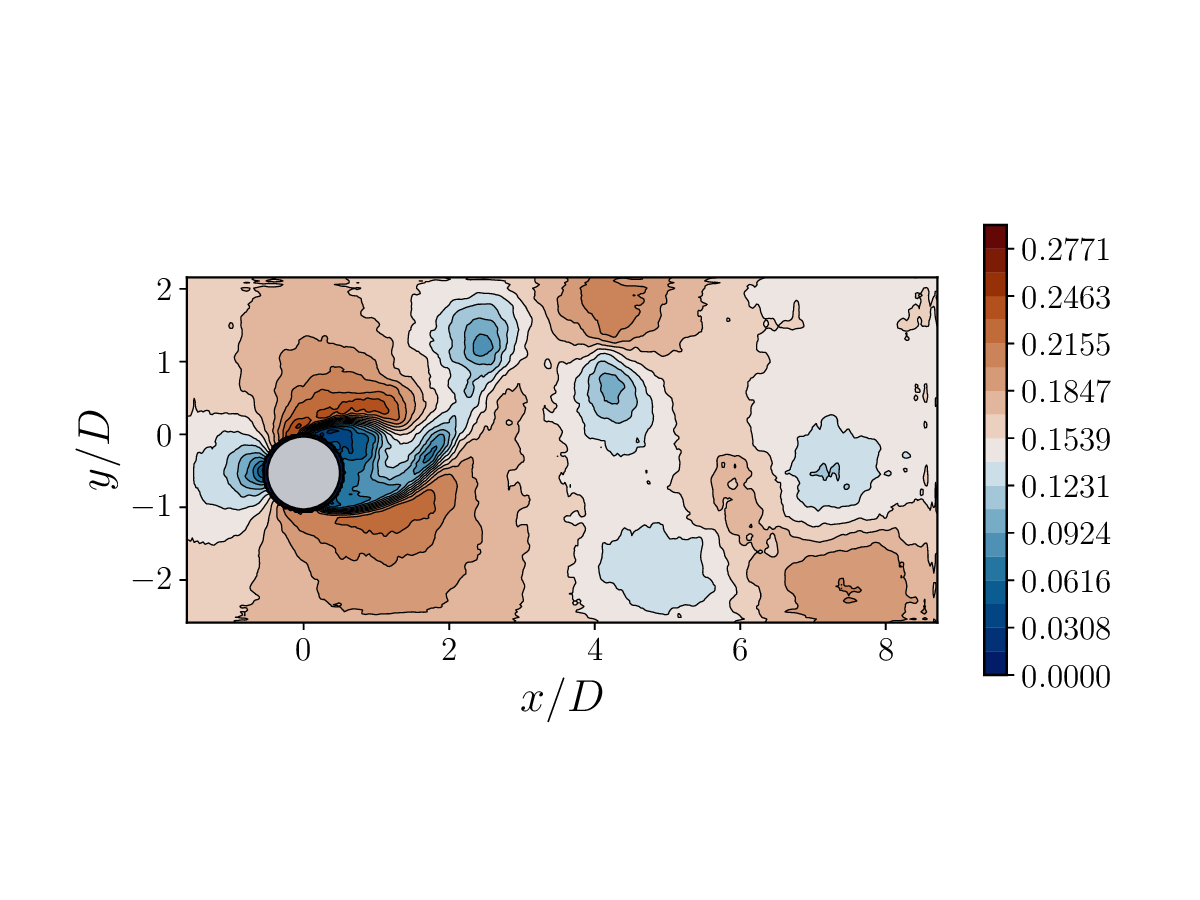}
    \caption{Phase-averaged velocity field comparison between ground truth (left) and VIVALDy predictions (right), colored by velocity magnitude $u_{\text{mag}} = \sqrt{u^2 + v^2}$. Results shown for upper branch ($U^* = 5.56$, top row) and lower branch ($U^* = 8.03$, bottom row) test cases. For each case, the depicted phase-average corresponds to the first bin.}
    \label{fig:04_3/phase_averaged_flow_fields}
\end{figure}

Figure~\ref{fig:04_3/phase_averaged_flow_fields} shows the computed phase-averaged fields for the upper and lower branch regimes analyzed in Section~\ref{sec:04_2}. Distinct wake topologies are clearly visible. The upper branch (top row) exhibits a broader wake with widely spaced vortex pairs, comparable to the 2Po vortex-shedding mode identified by Morse and Williamson \cite{morse2009prediction}. Instead, the lower branch (bottom row) presents a narrower wake with more closely spaced vortex structures, closely resembling the 2P vortex-shedding mode reported by Morse and Williamson \cite{morse2009prediction} and O'Neill \textit{et al.} \cite{o2021wake}.

VIVALDy's phase-averaged predictions (right column) closely align with ground truth (left column). The model reproduces both wake topologies with only minor deviations from target vortex patterns, preserving all dominant wake structures. Consequently, the NRMSE errors reported previously reflect finer flow structures, residual stochastic fluctuations not observed after the phase-averaging process, and measurement noise.

\subsubsection{Probability Density Function Visualization}

The flow statistical distribution visualization complements the quantitative Wasserstein distances results. Gaussian kernel density estimation was used to compute the probability density functions of the ground truth PIV snapshots and the VIVALDy predictions. 

\begin{figure}[h!]
    \centering
    \includegraphics[width=0.45\textwidth]{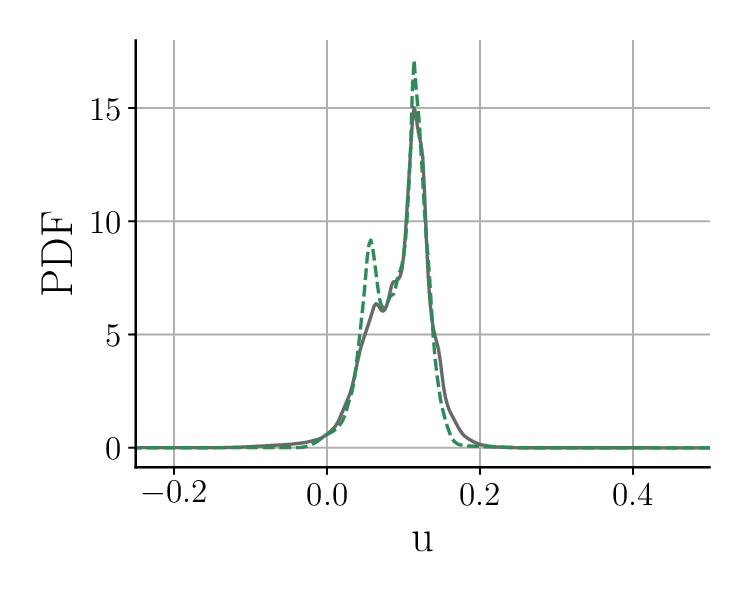}
    \includegraphics[width=0.45\textwidth]{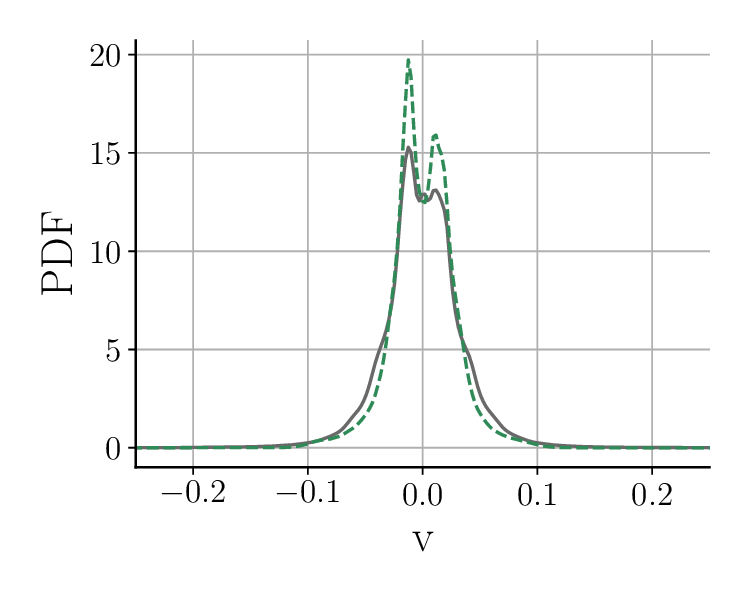}\\
    \vspace{3mm}
    \includegraphics[width=0.45\textwidth]{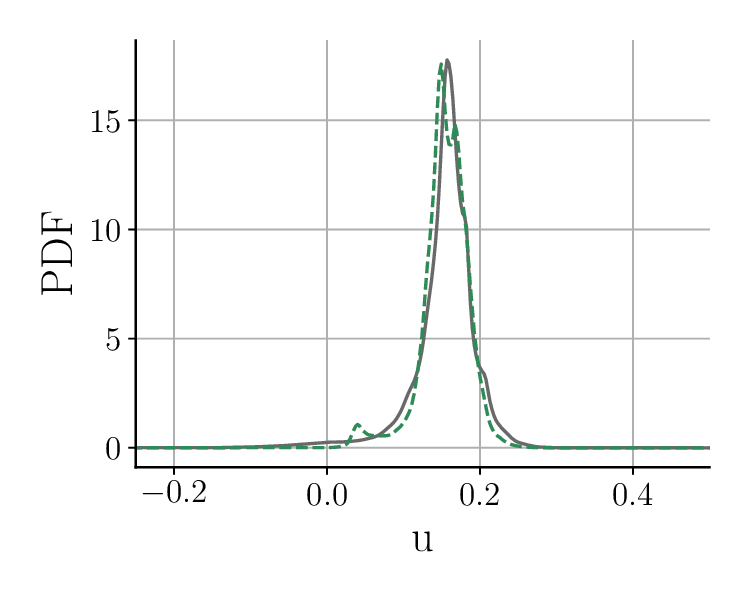}
    \includegraphics[width=0.45\textwidth]{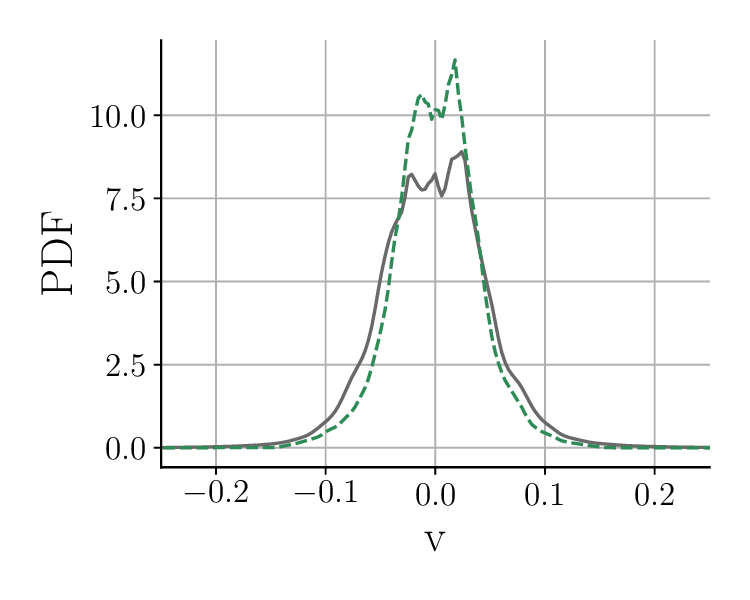}
    \caption{Probability density functions comparison between ground truth (gray) and VIVALDy predictions (green). Results shown for upper branch ($U^* = 5.56$, top row) and lower branch ($U^* = 8.03$, bottom row) test cases. For each case, the depicted phase-average corresponds to the first bin.}
    \label{fig:04_3/physical_space_pdfs}
\end{figure}

The computed PDFs for the upper and lower branch regimes are shown in Figure~\ref{fig:04_3/physical_space_pdfs}. The two operating conditions exhibit distinct velocity distributions, reflecting their different wake dynamics. Both cases show bimodal $v$-component distributions characteristic of alternating vortex shedding, while the $u$-component distributions differ significantly between regimes. The lower branch case, characterized by highly periodic oscillations, displays a symmetric $u$-component distribution. In contrast, the upper branch exhibits an asymmetric $u$-component distribution, reflecting its irregular dynamics.

VIVALDy's predictions show close shape preservation and reasonable alignment with the ground truth PDFs of both operating conditions distribution. However, some deviation are observed. In the upper branch $u$-component, a significant overestimation is observed in a secondary peak. Peak overestimation and lower variance are also observed for the $v$-component of both regimes, consistent with the higher Wasserstein distances found for this velocity component. The predictioin asymmetry quantified in the Wasserstein distance is also observable in the distribution, with the $u$-component predictions showing closer agreement with the ground truth than the $v$-component ones.

% ------------------------------ Discussion

\section{Discussion}
\label{sec:05_discussion}

\subsection{Rate-Distortion-Perception Trade-offs in Adversarial Training}

The empirical results of the ablation study (Section \ref{sec:04_1}) show that improvements in distributional alignment do not necessarily correspond to improvements in reconstruction accuracy. Indeed, the model trained with $\alpha=0.02$ showed lower Wasserstein distances but worse NRMSE compared to the benchmark. This apparent contradiction can be understood through information theory principles. Classical rate-distortion theory quantifies the fundamental trade-off between compression efficiency (rate) and reconstruction quality (distortion) \cite{shannon1959coding, cover2006elements}, however it does not account for statistical fidelity. Blau and Michaeli \cite{blau2019rethinking} extended this framework to include distributional alignment (termed perception quality) through the rate-distortion-perception trade-off. They demonstrate that requiring high distributional alignment generally elevates the rate-distortion curve, necessitating sacrifices in either compression rate or reconstruction accuracy.

The ablation study results align with this theoretical framework. Because the latent space dimension remains fixed at three, the compression rate is constant in this work. Consequently, increased distortion is expected and empirically observed for $\alpha = 0.02$. Conversely, with minimal adversarial weighting ($\alpha = 0.002$), slightly improved NRMSE values are observed for both velocity components, but at the cost of worse Wasserstein distances, consistent with the Blau and Michaeli framework. Notably, $\alpha=0.2$ appears to yield a more advantageous trade-off, achieving better Wasserstein distances for $v$ coupled with improved NRMSE for both components, despite slightly worse distributional alignment for the $u$ component. 

Interestingly, the $\alpha=0.2$ model's greatest improvement occurs in the Transition case. As reported in Section \ref{sec:02_dataset/2d_flow_snapshots}, this case appears exclusively in the test set. This improved performance on an unseen operating condition suggests that, by forcing the optimization to preserve the statistical properties of the flow field, adversarial training enhances model generalizability. Therefore, the $\alpha$ parameter enables the proposed hybrid $\beta$-VAE-GAN to achieve substantial improvements over the benchmark model while maintaining robust performance across diverse operating conditions.

\subsection{Limitations and Future Directions}
Throughout the latent and inference analyses, VIVALDy's predictions consistently exhibit narrower PDFs and latent space volume contractions compared to the experimental fields. This prediction behavior reflects several fundamental challenges facing the model.

The data preprocessing strategy, specifically the clipping of input fields at $\pm \sigma_c$, explicitly removes extreme outlier events from the training set, placing an upper bound on the learnable variance. Moreover, the chosen $\beta$-VAE-GAN architecture compress the data into just three latent variables, forcing the model to retain only dominant structures while discarding finer details. This drastic dimensionality reduction combines with two limitation imposed by the experimental dataset: limited training data and inherent measurement noise. While this work uses 900 snapshots per operating condition (10,800 total), comparable autoencoder-based studies on numerical simulations datasets typically access to $10^5$ samples \cite{solera2024beta, fukami2023grasping}. Other studies, access to data quantities comparable to this study but operates on single flow conditions  \cite{williams2023sensing}, whereas VIVALDy must generalize across diverse operating regimes. Given these constraints, the narrower PDFs can be interpreted as an implicit model regularization that prioritizes dominant dynamical structures, enhancing robustness at the expense of reproducing the full fluctuation spectrum. Indeed, VIVALDy demonstrates adequate accuracy across all relevant flow regimes and preserves the dominant wake topology in phase-averaged visualizations. In correlation analyses, the transformer successfully maintains variable relationships, \text{i.e} the underlying mode interactions, even as these relationships vary across flow configurations. This performance level is satisfactory for surrogate models intended for flow control or rapid design optimization applications.

Future iterations of the VIVALDy framework could address these limitations through two primary strategies. First, the dataset could be enriched with high-fidelity numerical simulations. A hybrid training approach, combining experimental and numerical data, would mitigate the impact of measurement noise and provide an extensive sample sizes. Second, the input space could be augmented with additional sensors. For instance, incorporating surface pressure probes would provide valuable information regarding the boundary layer state. This would enhance the model's predictive capability, particularly in regimes characterized by small cylinder displacements (such as the Initial and Asynchrony branches), where the current displacement-only signal provides limited observability of the flow state.

% ------------------------------- Conclusion

\section{Conclusions}
\label{sec:06_conclusions}

Reduced-order models are frameworks designed to create numerically efficient surrogate models of turbulent flows. Traditional ROM approaches face fundamental challenges in achieving computational efficiency while maintaining accuracy across diverse operating conditions. This paper introduces VIVALDy, a novel machine-learning framework that addresses these limitations through three innovations: masked convolutions to handle complex solid-fluid interfaces; a hybrid $\beta$-VAE-GAN architecture to learn informative and statistically consistent latent representations; and a bidirectional transformer to map minimal sensor inputs to the underlying flow dynamics. When validated on a vortex-induced vibration problem, VIVALDy successfully reconstructs the turbulent flow around an oscillating cylinder using only the body's displacement as input, demonstrating robust performance even in previously unseen regimes.

To demonstrate the framework's capabilities, VIVALDy is applied to a vortex-induced vibration problem: reconstructing the turbulent flow around an oscillating cylinder using only the body displacement as input measurement. This fluid-structure interaction problem represents a challenging test case due to moving  solid-fluid interface and irregularly changing wake dynamics, requiring generalization across operating regimes. The model is validated against an experimental dataset spanning diverse VIV conditions, demonstrating the general applicability of the proposed innovations.

At first, an ablation study was performed comparing three $\beta$-VAE-GAN configurations with different adversarial weight parameters $\alpha$ against a benchmark $\beta$-VAE. The study revealed fundamental trade-offs between reconstruction accuracy and distributional alignment, consistent with rate-distortion-perception theory. The optimal configuration ($\alpha=0.2$) enhanced model generalizability, particularly for unseen flow conditions. Using this optimal configuration, a detailed analysis of the latent space was performed on two test-set operating conditions. The results demonstrate that the encoded latent variables retain characteristic dynamical signatures of the flow, with mode pairs tracing annular orbits associated with vortex shedding oscillations in the wake. The analysis revealed nonlinear mode interactions not captured by linear decomposition methods, with anti-correlation patterns between latent variables aligning with literature evidence of competing vortex shedding modes. This suggests VIVALDy's latent space captures physically meaningful dynamics often missed by traditional linear ROM methods. The bidirectional transformer successfully predicts these attractor shapes and variable relationships using only cylinder displacement input. When integrated with the $\beta$-VAE-GAN decoder during inference, this approach delivers robust performance across all VIV regimes, accurately reproducing both coherent wake structures and statistical properties under diverse operating conditions. 

Beyond VIV applications, the hybrid architecture and masked convolution approach provide a general framework for fluid-dynamics problems involving complex geometries. The demonstrated ability to capture nonlinear mode interactions positions this approach as a valuable tool for understanding dynamics not fully captured by traditional linear methods. The successful application of adversarial training principles suggests broader potential for incorporating distribution-preserving objectives in scientific machine-learning, particularly where statistical fidelity is as important as reconstruction accuracy. Moreover, VIVALDy's single-sensor prediction capability across different flow states makes it versatile for diverse applications: as sparse sensing reconstruction \cite{fukami2021global,williams2023sensing,vishwasrao2025diff}, latent space data assimilation \cite{amendola2020dataassimilationlatentspace}, initialization of lagged observations for dynamic forecasting models such as transformers \cite{solera2024beta}, and development of optimal control strategies \cite{cremades2024identifying,beneitez2025improving}. Furthermore, the learned latent space topology offers geometric interpretation of flow dynamics, enabling the development of control strategies that guide dynamics along desired trajectories directly in the latent space \cite{fukami2024data}. VIVALDy thus represents an advancement toward sophisticated reduced-order models capable of mimicking physical systems, enabling practical flow control and optimization strategies.

\section*{Data availability}
The dataset used in this study is accessible at: \url{https://doi.org/10.57745/HPA87O}

\section*{Code availability}
The inference and training code, framework architectures, and trained weights are accessible at: \url{https://github.com/NiccoloTonioni/vivaldy_code}

\section*{Acknowledgments}
This work was supported by the French government program "Investissements d'Avenir" (EUR INTREE, reference ANR-18-EURE-0010 and LABEX INTERACTIFS, reference ANR-11-LABX-0017-01) and  was performed using HPC resources from GENCI–IDRIS (Grant 20XX-AD011015409). RV acknowledges the financial support from ERC grant no. 2021-CoG-101043998, DEEPCONTROL. Views
and opinions expressed are however those of the author(s) only and do not necessarily reflect those of the European Union or the European Research Council. Neither the European Union nor the granting authority can be held responsible for them.

\bibliographystyle{unsrtnat}
\bibliography{references}  %%% Uncomment this line and comment out the ``thebibliography'' section below to use the external .bib file (using bibtex) .




% 	\bibitem{keshet2016prediction}
% 	Keshet, Renato, Alina Maor, and George Kour.
% 	\newblock Prediction-Based, Prioritized Market-Share Insight Extraction.
% 	\newblock In {\em Advanced Data Mining and Applications (ADMA), 2016 12th International 
%                       Conference of}, pages 81--94,2016.

% \end{thebibliography}

\end{document}